\documentclass[journal]{IEEEtran}
\usepackage{cite}

\usepackage[dvips]{graphicx}
\usepackage{amsmath,amsfonts,amssymb,amsthm}
\usepackage{mathtools}
\usepackage{widetext} 
\usepackage{float}
\usepackage{comment}

\usepackage{stfloats}
\begin{document}
%
\title{Wavefront errors in two-wavelength\\ adaptive optics systems}
%
%
%

\author{Milo W. Hyde IV,~\IEEEmembership{Senior Member,~IEEE,}
        Matthew Kalensky,
        and~Mark F. Spencer~\IEEEmembership{Fellow,~SPIE}
\thanks{M. Hyde is with Epsilon C5I, Beavercreek, OH 45431 USA (e-mail: mhyde@epsilonsystems.com).}%
\thanks{M. Spencer is with the Office of the Under Secretary of Defense for Research and Engineering, Washington, DC 20301-3030 USA}
\thanks{Matthew Kalensky is with the Naval Surface Warfare Center Dahlgren
Division, Dahlgren, VA 22448 USA.}
\thanks{Manuscript received XXX XX, 2025; revised XXX XX, 2025.}}

%
%

\markboth{IEEE TRANSACTIONS ON ANTENNAS AND PROPAGATION}%
{Hyde~IV \MakeLowercase{\textit{et al.}}: Wavefront errors in two-wavelength AO sytems}
%


\newcommand{\Langle}{\left\langle}
\newcommand{\Rangle}{\right\rangle}
\newcommand{\eq}[1]{Eq.~(\ref{#1})}
\newcommand{\eqs}[2]{Eqs.~(\ref{#1}) and~(\ref{#2})}
\newcommand{\fig}[1]{Fig.~\ref{#1}}
\newcommand{\figs}[2]{Figs.~\ref{#1} and~\ref{#2}}
\newcommand{\uvect}[1]{$\hat{\boldsymbol{#1}}$}
\newcommand{\uvecm}[1]{\hat{\boldsymbol{#1}}}
\newcommand{\GF}[1]{\Gamma\left({#1}\right)}

\maketitle

\begin{abstract}
Two-wavelength adaptive optics (AO) systems sense turbulence-induced wavefront distortions using an artificial beacon or natural guidestar at one wavelength, while correcting and possibly transmitting at another.  Although most existing AO systems employ this methodology, the literature on atmospheric turbulence correction and AO system design generally focuses on performance at a single wavelength, neglecting the two-wavelength nature of the problem.  In this paper, we undertake a rigorous study of the relevant wavefront errors necessary to quantify two-wavelength AO system performance.  

Since most AO systems employ separate tilt and higher-order correcting subsystems, our analysis mirrors this division, and we begin with higher-order wavefront errors.  Utilizing Mellin transform techniques, we derive closed-form relations for the piston-removed and piston- and tilt-removed variances.  The former is a measure of the total, residual wavefront error that a two-wavelength AO systems experiences; while the latter, quantifies the residual wavefront error due to higher-order aberrations.  

We then proceed to tilt or tracking errors and derive the two-wavelength Zernike- and gradient-tilt variances.  Zernike tilt is the actual amount of tilt in the turbulent atmosphere; yet, most AO tracking subsystems measure gradient tilt.  Consequently, we also derive the two-wavelength gradient-tilt, Zernike-tilt variance---also known as centroid anisoplanatism---to quantify this error. 

Lastly, we validate our analysis by performing two-wavelength wave-optics simulations and comparing the results to theory.  We observe excellent agreement among the simulated results and our theoretical predictions. 

The analysis and findings presented in this paper will be useful in the characterization of existing, and the design of new, two-wavelength AO systems.
\end{abstract}


%
\IEEEpeerreviewmaketitle

\section{Introduction}\label{sec:intro}
Adaptive optics (AO) provides a means to sense and correct for wavefront aberrations that accumulate as light propagates through a random or turbulent medium.  With that said, astronomical and power-beaming telescopes often employ two-wavelength AO. Here, practitioners perform the sensing at one wavelength $\lambda_\text{B}$ and correction at another wavelength $\lambda_\text{T}$. For practical reasons, they typically associate $\lambda_\text{B}$ with the light from a beacon in the form of a natural or laser guidestar and $\lambda_\text{T}$ with the light transmitted through the atmosphere to focus an image or project a beam.  Despite the ubiquitous nature of this setup, the published AO textbooks~\cite{Hardy,Roddier,Perram,PAO,Merritt} do not typically mention the two-wavelength nature of the problem when formulating error budgets at the system level.

When dealing with atmospheric turbulence, the various formulations for the two-wavelength wavefront errors often result in integral expressions. Such expressions are often difficult to simplify into error budgets at the system level. As pointed out by Fugate~\emph{et al.}~\cite{Fugate:23} in their tribute to David L.~Fried, scientists and engineers prefer to use closed-form expressions because of the physical insight they provide.  

Nonetheless, several researchers have endeavored to quantify two-wavelength wavefront errors. For example, Hogge and Butts~\cite{Hogge:82} were the first to develop an integral expression for the two-wavelength, optical-path-difference (OPD) variance.  Despite being published in 1982, this foundational paper has only been cited 15 times (according to the publisher), which speaks to our earlier point regarding scientists and engineers' affinity for closed-form expressions.  

To formulate their integral expression, Hogge and Butts assumed that the effects of scintillation were negligible. Using the Hufnagel--Valley model for the index of refraction structure constant $C_n^2$~\cite{Sasiela:07}, they numerically evaluated their integral expression and found that as long as $\lambda_\text{B} < \lambda_\text{T}$ and $|\lambda_\text{B}-\lambda_\text{T}|$ was less than a few microns, the two-wavelength OPD variance did not significantly degrade system performance. Because of the weak-scintillation assumption and the $C_n^2$ model, Hogge and Butts' findings were more applicable to up-looking, astronomical and power-beaming scenarios and are likely the reasons why the published AO textbooks~\cite{Hardy,Roddier,Perram,PAO,Merritt} to date, which are heavily focused on astronomical AO, do not formulate two-wavelength error budgets.

Hyde~\emph{et al.}~\cite{10529268} recently built on the foundational work of Hogge and Butts, but used a path-invariant model for $C_n^2$.  In so doing, their findings are more applicable to horizontal-looking, remote-sensing, and directed-energy scenarios. Using Mellin transform techniques, they were able to derive a closed-form expression for the plane-wave, two-wavelength OPD variance. This simple expression resulted in two physical insights that were missing from previous formulations: The two-wavelength wavefront error is (1) weakly dependent on the aperture diameter and (2) approximately wavelength shift-invariant (i.e., it depends only on the difference between $\lambda_\text{B}$ and $\lambda_\text{T}$ and not on the wavelengths themselves). Even though (1) and (2) offer new physical insights that scientists and engineers can now leverage, there is more that can be gleaned.

In this paper, we develop new formulas consistent with the error budgets formulated in published AO textbooks~\cite{Hardy,Roddier,Perram,PAO,Merritt}.  In particular, we derive closed-form expressions for the spherical-wave, two-wavelength tilt and higher-order wavefront errors.  These expressions build upon the developments of Hogge and Butts~\cite{Hogge:82} and Hyde~\emph{et al.}~\cite{10529268} and again use a path-invariant model for $C_n^2$. Thus, these expressions are more applicable to horizontal paths and artificial beacon AO systems (for more insight, see Refs.~\cite{Fried:98,Barchers:2003,Venema:08_2,Steinbock:14,Banet:20,Spencer:21,Spencer:22,Hyde:2024,Kalensky:24}).

One reason for formulating error budgets in terms of tilt and higher-order wavefront errors is that AO systems typically use separate subsystems to sense and correct these aberrations.  By decomposing the two-wavelength wavefront error into tilt and higher-order terms, we can better account for performance at the system level.

With the above decomposition in mind, great care must be taken in formulating the problem at hand.  We must understand the higher-order wavefront error in terms of orthogonal Zernike polynomials. In so doing, we formulate the piston-removed and piston- and tilt-removed versions of the two-wavelength wavefront error. The former is necessary since higher-order wavefront sensors, like the centroid-based Shack--Hartmann wavefront sensor, do not sense piston.  The latter is then necessary when accounting for tilt compensation from a separate subsystem. What is more, the Zernike-mode decomposition of the problem allows for future efforts to account for higher-order compensation in a straightforward way~\cite{Tyson:82}.

In addition, we must also understand the various forms of tilt, including Zernike tilt or ``Z-tilt'' and gradient tilt or ``G-tilt.''  For all intents and purposes, Z-tilt is what practitioners would like to sense and correct using an AO system.  Noll~\cite{Noll:76} was the first to show that Z-tilt comprises 87\% of the total wavefront error (at one wavelength). In the absence of irradiance fluctuations (e.g., noise~\cite{Tyler:82}, speckle~\cite{Burrell:23}, and scintillation~\cite{Mitchell:25}), a centroid-based tilt sensor or ``tracker'' within an AO system senses G-tilt, not Z-tilt.  This difference leads to G-tilt, Z-tilt error, which is often referred to as centroid anisoplanatism~\cite{Yura:85}.

In what follows, we derive closed-form expressions for all of these two-wavelength wavefront errors or variances. In Section~\ref{sec:thy}, we formulate the higher-order wavefront errors in terms of Zernike polynomials and derive piston-removed and piston- and tilt-removed errors consistent with past formulations at one wavelength. In Section~\ref{sec:tlt}, we focus on tilt errors and derive expressions for the Z-tilt, G-tilt, and G-tilt, Z-tilt angle variances.  We validate our theory in Section~\ref{sec:sim} by comparing results from two-wavelength wave-optics simulations to our tilt and higher-order wavefront error expressions.  Lastly, we conclude this paper with a short summary of our work and contributions.

\section{Higher-order wavefront errors}\label{sec:thy}
In this section, we formulate closed-form expressions for two-wavelength higher-order wavefront errors, specifically, the two-wavelength, Zernike-mode optical path difference (OPD), the piston-removed OPD, and the piston- and tilt-removed OPD variances. Recall that in the pupil plane, we can relate the phase $\phi$ (at the transmit wavelength $\lambda_\text{T}$) to the OPD $\Delta\ell$ using the following relationship: 
\begin{equation}\label{eq:WF_phase_relation}
\phi\left(\lambda_\text{T}\right) = k_\text{T} \Delta \ell,
\end{equation}
where $k_\text{T}=2\pi/\lambda_\text{T}$ is the wavenumber. Equation~\eqref{eq:WF_phase_relation} enables the reader to make the following closed-form OPD expressions consistent with the various error-budget formulations found in AO textbooks~\cite{Hardy,Roddier,Perram,PAO,Merritt}.

\subsection{Two-wavelength Zernike-mode OPD variance}\label{sec:thy1}
Let us start with the optical path length (OPL) at wavelength $\lambda$ expanded in terms of Zernike polynomials, namely,
\begin{equation}\label{eq:Z1}
\ell\left(\boldsymbol{\rho},\lambda\right) = \sum\limits_{m=0}^{\infty} a_m\left(\lambda\right) Z_m\left(\frac{\rho}{D/2},\phi\right),
\end{equation}
where $a_m$ is the weight of the Zernike polynomial $Z_m$ in meters, $m$ is a single index corresponding to a unique double index $i,j$ that specifies the radial and azimuthal variation of $Z_m$~\cite{Noll:76,OIAWA,Lakshminarayanan10042011,Sasiela:07}, and $D$ is the diameter of the receiving aperture.  The OPD at two wavelengths is clearly
\begin{equation}\label{eq:Z2}
\begin{gathered}
\ell\left(\boldsymbol{\rho},\lambda_\text{B}\right) - \ell\left(\boldsymbol{\rho},\lambda_\text{T}\right) = \Delta \ell\left(\boldsymbol{\rho}\right) \hfill \\
\quad = \sum\limits_{m=0}^{\infty}\left[ a_m\left(\lambda_\text{B}\right) - a_m\left(\lambda_\text{T}\right) \right] Z_m\left(\frac{\rho}{D/2},\phi\right). \hfill
\end{gathered}
\end{equation}
We are ultimately interested in the mean (over turbulence realizations) of the square of \eq{eq:Z2} spatially averaged over $D$, that is,
\begin{equation}\label{eq:Z3}
\Langle \Delta \ell^2\Rangle = \frac{1}{A}\iint_{-\infty}^{\infty} \operatorname{circ}\left(\frac{\rho}{D}\right) \Langle\left[\ell\left(\boldsymbol{\rho},\lambda_\text{B}\right) - \ell\left(\boldsymbol{\rho},\lambda_\text{T}\right) \right]^2 \Rangle\text{d}^2 \rho,    
\end{equation}
where $A = \pi\left(D/2\right)^2$.  Substituting the right-hand side of \eq{eq:Z2} into \eq{eq:Z3} and expanding produces
\begin{equation}\label{eq:Z4}
\begin{gathered}
\Langle \Delta \ell^2\Rangle = \sum\limits_{m=0}^\infty \sum\limits_{n=0}^\infty \left[ \Langle a_n\left(\lambda_\text{B}\right) a_m\left(\lambda_\text{B}\right) \Rangle \right. \hfill \\
\left. \; +\, \Langle a_n\left(\lambda_\text{T}\right) a_m\left(\lambda_\text{T}\right) \Rangle  - 2 \Langle a_n\left(\lambda_\text{B}\right) a_m\left(\lambda_\text{T}\right) \Rangle \right] \hfill \\
\;\times \frac{1}{A} \iint_{-\infty}^{\infty} \operatorname{circ}\left(\frac{\rho}{D}\right) Z_n\left(\frac{\rho}{D/2},\phi\right) Z_m\left(\frac{\rho}{D/2},\phi\right) \text{d}^2 \rho. \hfill 
\end{gathered}
\end{equation}
The double integral over $\boldsymbol{\rho}$ is the Zernike polynomial orthogonality relation and equals the Kronecker delta $\delta_{nm}$~\cite{Noll:76,OIAWA,Lakshminarayanan10042011}.  This trivially eliminates one of the sums yielding
\begin{equation}\label{eq:Z5}
\begin{gathered}
\Langle \Delta \ell^2\Rangle = \sum\limits_{m=0}^\infty \Langle a_m^2\left(\lambda_\text{B}\right)  \Rangle +  \sum\limits_{m=0}^\infty \Langle a_m^2\left(\lambda_\text{T}\right)  \Rangle \hfill \\
\quad -\, 2  \sum\limits_{m=0}^\infty  \Langle a_m\left(\lambda_\text{B}\right) a_m\left(\lambda_\text{T}\right) \Rangle. \hfill 
\end{gathered}
\end{equation}
By inspection, we can conclude that the OPD variance of a Zernike mode is therefore 
\begin{equation}\label{eq:Z6}
\Langle \Delta \ell_m^2\Rangle =  \Langle a_m^2\left(\lambda_\text{B}\right)  \Rangle +    \Langle a_m^2\left(\lambda_\text{T}\right)  \Rangle - 2 \Langle a_m\left(\lambda_\text{B}\right) a_m\left(\lambda_\text{T}\right) \Rangle.
\end{equation}
We now focus on evaluating the last (covariance) term in \eq{eq:Z6}, since the others can be found from it by setting $\lambda_\text{B}=\lambda_\text{T}$.

\subsection{Two-wavelength Zernike-mode covariance function}\label{sec:thy2}
To evaluate the Zernike-mode covariance term in \eq{eq:Z6}, we need an expression for $a_m$.  This can be derived quite easily using the Zernike polynomial orthogonality relation, namely,
\begin{equation}\label{eq:Z7}
a_m\left(\lambda\right) = \frac{1}{A}\iint_{-\infty}^{\infty} \operatorname{circ}\left(\frac{\rho}{D}\right) \ell\left(\boldsymbol{\rho},\lambda\right)  Z_m\left(\frac{\rho}{D/2},\phi\right) \text{d}^2 \rho.     
\end{equation}
Substituting the Fourier transform of a Zernike polynomial~\cite{Noll:76,Sasiela:07,Lakshminarayanan10042011} into \eq{eq:Z7} and rearranging the integrals produces
\begin{equation}
\begin{gathered}
 a_m\left(\lambda\right) = \frac{\pi}{A} \iint_{-\infty}^{\infty} Q_m\left(\boldsymbol{f}\right) \hfill \\
 \quad \times \iint_{-\infty}^{\infty}\ell\left(\boldsymbol{\rho},\lambda\right) \exp\left(-\text{j} 2\pi \boldsymbol{f}\cdot \frac{\boldsymbol{\rho}}{D/2}\right) \text{d}^2 \rho \text{d}^2 f. \hfill
\end{gathered}
\end{equation}

We now compute the covariance and obtain
\begin{equation}\label{eq:Z9}
\begin{gathered}
\Langle a_m\left(\lambda_\text{B}\right) a_m\left(\lambda_\text{T}\right) \Rangle = \frac{\pi^2}{A^2} \iiiint_{-\infty}^{\infty} Q_m\left(\boldsymbol{f}_1\right) Q_m^*\left(\boldsymbol{f}_2\right)  \hfill \\
\; \times \iiiint_{-\infty}^{\infty}  \Langle \ell\left(\boldsymbol{\rho}_1,\lambda_\text{B}\right) \ell\left(\boldsymbol{\rho}_2,\lambda_\text{T}\right) \Rangle  \exp\left(-\text{j} 2\pi \boldsymbol{f}_1\cdot \frac{\boldsymbol{\rho}_1}{D/2}\right) \hfill \\
\; \times \exp\left(\text{j} 2\pi \boldsymbol{f}_2\cdot \frac{\boldsymbol{\rho}_2}{D/2}\right)  \text{d}^2 \rho_1 \text{d}^2 \rho_2 \text{d}^2 f_1 \text{d}^2 f_2. \hfill 
\end{gathered}
\end{equation}
The statistical moment on the second line of \eq{eq:Z9} is the two-wavelength OPL covariance, which is equal to
\begin{equation}\label{eq:Z10}
B_\ell\left(\boldsymbol{\rho}_1-\boldsymbol{\rho}_2,\lambda_\text{B},\lambda_\text{T}\right) = \frac{1}{k_\text{B} k_\text{T}} B_S\left(\boldsymbol{\rho}_1-\boldsymbol{\rho}_2,\lambda_\text{B},\lambda_\text{T}\right),
\end{equation}
where $B_S$ is the two-wavelength phase covariance function~\cite{1140133,Ishimaru:99}, i.e.,
\begin{equation}\label{eq:Z11}
\begin{gathered}
B_S\left(\rho,\lambda_\text{B},\lambda_\text{T}\right) = 4 \pi^2 k_\text{B} k_\text{T} \int_0^z \int_0^\infty \kappa \Phi_n\left(\kappa,\zeta\right) \hfill \\ \; \times J_0\left(\frac{\zeta}{z} \kappa \rho\right)\cos\left[\frac{z}{2k_\text{B}}\frac{\zeta}{z}\left(1-\frac{\zeta}{z}\right)\kappa^2\right] \hfill \\
\; \times \cos\left[\frac{z}{2k_\text{T}}\frac{\zeta}{z}\left(1-\frac{\zeta}{z}\right)\kappa^2\right] \text{d}\zeta \text{d}\kappa, \hfill
\end{gathered}    
\end{equation}
and $\Phi_n$ is the index of refraction power spectrum.  Note that $\Phi_n$ only depends on $\zeta$ (propagation distance in the turbulent medium) via the index of refraction structure constant $C_n^2$~\cite{Andrews:05,Sasiela:07,Ishimaru:99,Tatarskii:61}.

Inserting \eqs{eq:Z10}{eq:Z11} into \eq{eq:Z9}, making the variable substitutions $\boldsymbol{\rho}' = \boldsymbol{\rho}_1$ and $\boldsymbol{\rho} = \boldsymbol{\rho}_1-\boldsymbol{\rho}_2$, and converting to polar coordinates yields
\begin{widetext}
\begin{equation}\label{eq:Z12}
\begin{gathered}
\Langle a_m\left(\lambda_\text{B}\right) a_m\left(\lambda_\text{T}\right) \Rangle = \frac{8\pi^4}{A} \iint_{-\infty}^{\infty} \left|Q_m\left(\boldsymbol{f}\right)\right|^2 \int_0^z \int_0^\infty \kappa \Phi_n\left(\kappa,\zeta\right) \cos\left[\frac{z}{2k_\text{B}}\frac{\zeta}{z}\left(1-\frac{\zeta}{z}\right)\kappa^2\right] \cos\left[\frac{z}{2k_\text{T}}\frac{\zeta}{z}\left(1-\frac{\zeta}{z}\right)\kappa^2\right] \hfill \\
\quad \times \int_0^\infty \rho J_0\left(\frac{\zeta}{z} \kappa \rho\right) J_0\left(2 \pi f \frac{\rho}{D/2}\right) \text{d}\rho \text{d}\kappa \text{d}\zeta \text{d}^2 f. \hfill
\end{gathered}
\end{equation}
With variable substitutions, the $\rho$ integral is the Bessel function orthogonality relation~\cite{Abramowitz1964,Gradshteyn2000} simplifying \eq{eq:Z12} to
\begin{equation}\label{eq:Z13}
\begin{gathered}
 \Langle a_m\left(\lambda_\text{B}\right) a_m\left(\lambda_\text{T}\right) \Rangle =  \frac{8\pi^2}{D} \int_{0}^z \int_0^\infty \left(\frac{\zeta}{z}\right)^{-1} \Phi_n\left(\kappa,\zeta\right) \cos\left[\frac{z}{2k_\text{B}}\frac{\zeta}{z}\left(1-\frac{\zeta}{z}\right)\kappa^2\right] \cos\left[\frac{z}{2k_\text{T}}\frac{\zeta}{z}\left(1-\frac{\zeta}{z}\right)\kappa^2\right] \hfill \\
\quad \times \iint_{-\infty}^{\infty} \left|Q_m\left(\boldsymbol{f}\right)\right|^2 \delta\left(f-\frac{1}{2\pi}\frac{D}{2}\frac{\zeta}{z}\kappa\right) \text{d}^2 f \text{d} \kappa \text{d} \zeta. \hfill      
\end{gathered}    
\end{equation}
Making use of Refs.~\cite{Noll:76, Sasiela:07}, the integrals over $\boldsymbol{f}$ are easy to compute and equal to
\begin{equation}\label{eq:Z14}
\begin{gathered}
\iint_{-\infty}^{\infty} \left|Q_m\left(\boldsymbol{f}\right)\right|^2 \delta\left(f-\frac{1}{2\pi}\frac{D}{2} \frac{\zeta}{z}\kappa\right) \text{d}^2 f= 4 \left(i+1\right) \frac{J_{i+1}^2\left[\kappa D \zeta/\left(2z\right)\right]}{\kappa D \zeta/\left(2z\right)},  
\end{gathered}
\end{equation}
where $i$ is the Zernike polynomial radial index corresponding to the single index $m$.  Since there is no longer a need to consider the azimuthal index $j$, we replace $m$ with $i$ hereafter.  

By substituting \eq{eq:Z14} into \eq{eq:Z13}, we arrive at the two-wavelength Zernike-mode covariance:
\begin{equation}\label{eq:Z15}
\begin{gathered}
\Langle a_i\left(\lambda_\text{B}\right) a_i\left(\lambda_\text{T}\right) \Rangle = \frac{64\pi^2}{D^2}\left(i+1\right) \int_{0}^z \left(\frac{\zeta}{z}\right)^{-2} \int_0^\infty \kappa^{-1} \Phi_n\left(\kappa,\zeta\right) \hfill \\
\quad \times \cos\left[\frac{z}{2k_\text{B}}\frac{\zeta}{z}\left(1-\frac{\zeta}{z}\right)\kappa^2\right] \cos\left[\frac{z}{2k_\text{T}}\frac{\zeta}{z}\left(1-\frac{\zeta}{z}\right)\kappa^2\right] J_{i+1}^2\left(\frac{D}{2}\frac{\zeta}{z}\kappa\right) \text{d}\kappa \text{d}\zeta. \hfill
\end{gathered}
\end{equation}
Finally, returning to \eq{eq:Z6} and inserting \eq{eq:Z15} yields the two-wavelength Zernike-mode OPD variance, which we can express concisely as
\begin{equation}\label{eq:Z16}
\begin{gathered}
\Langle \Delta \ell_i^2\Rangle = \frac{64\pi^2}{D^2}\left(i+1\right) \sum\limits_{k=1}^5 c_k \int_{0}^z \left(\frac{\zeta}{z}\right)^{-2} \int_0^\infty \kappa^{-1} \Phi_n\left(\kappa,\zeta\right)  J_{i+1}^2\left(\frac{D}{2}\frac{\zeta}{z}\kappa\right) \cos\left[\alpha_k\frac{\zeta}{z}\left(1-\frac{\zeta}{z}\right)\kappa^2\right] \text{d}\kappa \text{d}\zeta, \hfill
\end{gathered}    
\end{equation}
where $\boldsymbol{c} = \begin{bmatrix} 1 & 1/2 & 1/2 & -1 & -1\end{bmatrix}$ and 
\begin{equation}\label{eq:Z17}
    \boldsymbol{\alpha} = \begin{bmatrix} 0 & \dfrac{z}{k_\text{T}} & \dfrac{z}{k_\text{B}} & z\dfrac{k_\text{T}-k_\text{B}}{2k_\text{T}k_\text{B}} & z\dfrac{k_\text{T}+k_\text{B}}{2k_\text{T}k_\text{B}}\end{bmatrix}.
\end{equation}

\subsection{Evaluating \eq{eq:Z16} using Mellin transforms}\label{sec:thy3}
We now evaluate \eq{eq:Z16} using Mellin transform techniques.  Substituting in the Kolmogorov power spectrum $\Phi_n$~\cite{Andrews:05,Sasiela:07,Ishimaru:99,Tatarskii:61} and simplifying produces
\begin{equation}\label{eq:Z18}
\begin{gathered}
\Langle \Delta \ell_i^2\Rangle = 2^{14/3} \frac{5}{9} \Gamma\begin{bmatrix}5/6\\2/3\end{bmatrix} 
 \sqrt{\pi} \left(i+1\right) D^{-2} \sum\limits_{k=1}^5 c_k \int_{0}^z C_n^2\left(\zeta\right) \left(\frac{\zeta}{z}\right)^{-2} \hfill \\
\quad \times \int_0^\infty \kappa^{-14/3}  J_{i+1}^2\left(\frac{D}{2}\frac{\zeta}{z}\kappa\right) \cos\left[\alpha_k\frac{\zeta}{z}\left(1-\frac{\zeta}{z}\right)\kappa^2\right] \text{d}\kappa \text{d}\zeta, \hfill    
\end{gathered}    
\end{equation}
where~\cite{Sasiela:07}
\begin{equation}
\Gamma\begin{bmatrix} a_1,a_2,\cdots,a_m\\b_1,b_2,\cdots,b_n\end{bmatrix} = \frac{\GF{a_1} \GF{a_2} \cdots \GF{a_m}}{\GF{b_1} \GF{b_2} \cdots \GF{b_n}}.
\end{equation}
Utilizing the Mellin convolution formula~\cite{Sasiela:07,Brychkov2018}, the $\kappa$ integral in \eq{eq:Z18} can be written as a contour integral such that
\begin{equation}\label{eq:Z19}
    \begin{gathered}
     \Langle \Delta \ell_i^2\Rangle = 2^{-31/6} \frac{5}{9} \Gamma\begin{bmatrix}5/6\\2/3\end{bmatrix} \sqrt{\pi} \left(i+1\right) D^{5/3} \sum\limits_{k=1}^5 c_k \frac{1}{\text{j}2\pi} \int_C \left(\frac{D^4}{64 \alpha_k^2}\right)^{-s} \Gamma\begin{bmatrix}s+i/2-5/12,s+i/2+1/12\\ -s+i/2+23/12,-s+i/2+29/12\end{bmatrix} \hfill \\ \quad
     \times\, \Gamma\begin{bmatrix}-s+7/6,-s+5/3,-s\\ -s+17/12,-s+23/12,s+1/2\end{bmatrix} \int_{0}^z C_n^2\left(\zeta\right) \left(\frac{\zeta}{z}\right)^{5/3-2s}\left(1-\frac{\zeta}{z}\right)^{2s}\text{d}\zeta \text{d}s, \hfill
    \end{gathered}
\end{equation}    
where the contour $C$ crosses the real $s$ axis between $-i/2+5/12 < \operatorname{Re}\left(s\right) < 0$ with $i \geq 1$.

To proceed further, we assume that $C_n^2$ is constant.  This approximation is generally applicable to horizontal propagation paths, which is germane here.  The remaining integral over $\zeta$ is equal to a beta function~\cite{Abramowitz1964, Gradshteyn2000}, which can be expressed in terms of gamma functions, namely,
\begin{equation}\label{eq:Z20}
\begin{gathered}
\int_{0}^z C_n^2\left(\zeta\right) \left(\frac{\zeta}{z}\right)^{5/3-2s}\left(1-\frac{\zeta}{z}\right)^{2s}\text{d}\zeta  = C_n^2 z \frac{2^{5/3}}{\pi}\Gamma\begin{bmatrix}s+1/2,s+1,-s+4/3,-s+11/6\\11/3\end{bmatrix}. 
\end{gathered}    
\end{equation}
Substituting \eq{eq:Z20} into \eq{eq:Z19} produces
\begin{equation}\label{eq:Z21}
    \begin{gathered}
        \Langle \Delta \ell_i^2\Rangle = \frac{2^{-7/2}}{\sqrt{\pi}} \frac{5}{9} \Gamma\begin{bmatrix} 5/6\\2/3,11/3\end{bmatrix} C_n^2 z \left(i+1\right) D^{5/3} \sum\limits_{k=1}^5 c_k \frac{1}{\text{j}2\pi}
        \int_C \left(\frac{D^4}{64 \alpha_k^2}\right)^{-s} \Gamma\begin{bmatrix}i/2-5/12+s,i/2+1/12+s,1+s\end{bmatrix} \hfill \\ \qquad
        \times\, \Gamma\begin{bmatrix}7/6-s,5/3-s,-s,4/3-s,11/6-s\\i/2+23/12-s,i/2+29/12-s,17/12-s,23/12-s\end{bmatrix}\text{d}s \hfill \\        
        \;=\frac{2^{-7/2}}{\sqrt{\pi}} \frac{5}{9}\Gamma\begin{bmatrix} 5/6\\2/3,11/3\end{bmatrix}C_n^2 z \left(i+1\right) D^{5/3} \sum\limits_{k=1}^5 c_k G^{3,5}_{5,7}\left(\frac{D^4}{64 \alpha_k^2}\left|\begin{matrix*}[c] -\dfrac{1}{6}, -\dfrac{2}{3},1,-\dfrac{1}{3},-\dfrac{5}{6}; \text{---}\text{---} \\[10pt] \dfrac{i}{2}-\dfrac{5}{12}, \dfrac{i}{2}+\dfrac{1}{12},1;-\dfrac{i}{2}-\dfrac{11}{12},-\dfrac{i}{2}-\dfrac{17}{12}, -\dfrac{5}{12},-\dfrac{11}{12}\end{matrix*}\right.\right), 
    \end{gathered}
\end{equation}
where $G$ is a Meijer G-function~\cite{Wolfram,Luke1975,Sasiela:07,Gradshteyn2000,Brychkov2018}.  Numerical routines to evaluate Meijer G-functions are available in MATLAB, Mathematica, and Python.  Note that there are several different definitions of a Meijer G-function.  Here, we use the one given in Ref.~\cite{Wolfram}, which is consistent with the Meijer G-function routines in Mathematica and Python.  \end{widetext}

\subsection{Asymptotic solution}\label{sec:asymZ}
Although we have obtained a closed-form answer for the two-wavelength Zernike-mode OPD variance, the result provides little insight into how $\Langle \Delta \ell_i^2\Rangle$ behaves versus $D$, $z$, $\lambda_\text{B}$-$\lambda_\text{T}$ separation, or Zernike-mode index $i$.  In addition, inspection of the contour integral in \eq{eq:Z21} reveals that it converges for all values of the argument $D^4/\left(64 \alpha_k^2 \right)$ when $C$ is closed to the left, encircling the poles at $s = -m - i/2+5/12,\, -m-i/2-1/12$, and $-m-1$ for $m = 0,\, 1,\, 2,\, \cdots$.  Convergence of the resulting sums, and therefore convergence of the Meijer G-function, is very slow for large values of $D^4/\left(64\alpha_k^2\right)$:
\begin{equation}
\left(\frac{D^2}{8\alpha_k}\right)^2 \sim \left[\frac{D^2}{8\left(z/k\right)^2}\right]^2 = \left[\frac{\pi\left(D/2\right)^2}{\lambda z}\right]^2,
\end{equation}
which is clearly related to the Fresnel number $N_F$~\cite{IFO,Gaskill1978}.  Most beam projection systems operate with $N_F > 1$, so large values of the argument are expected.  Therefore, we seek an asymptotic solution to the integral in \eq{eq:Z21}.

We obtain such a solution by including contributions from poles to the right of $C$, i.e., $s=m,\, m+7/6,\, m+4/3,\, m+5/3$, and $m+11/6$ for $m = 0,\, 1,\, 2, \cdots, M-1$.  Applying Cauchy's integral formula~\cite{Arfken2013,Gbur_2011} and proceeding in ascending order, the contributions from the poles at $s = 0$ and $s = 1$ are zero. The dominant contribution comes from the pole at $s = 7/6$ and $\Langle \Delta \ell_i^2 \Rangle$ is approximately
\begin{equation}\label{eq:Z23}
\begin{gathered}
\Langle \Delta \ell_i^2 \Rangle \approx {2^{-14/3}} \frac{3}{\pi^{7/3}}
\Gamma\begin{bmatrix}-7/6,13/6\\2/3\end{bmatrix} C_n^2 z^{10/3} D^{-3} \left(i+1\right)\hfill \\
 \;\times \left[2^{4/3}\left(\lambda_\text{T}^{7/3}+\lambda_\text{B}^{7/3}\right) - \left|\lambda_\text{T}-\lambda_\text{B}\right|^{7/3} - \left(\lambda_\text{T}+\lambda_\text{B}\right)^{7/3}\right]. \hfill
\end{gathered}    
\end{equation}
This approximation improves the larger the $N_F$.  Including more pole contributions results in a more accurate approximation up to a certain point, and then the solution diverges (see Appendix~\ref{appC} for the asymptotic series). The requisite number of terms depends in a complex way on $D$, $i$, $z$, $\lambda_\text{B}$, and $\lambda_\text{T}$, but can be found rather easily through trial and error. 

\subsection{Two-wavelength piston-removed OPD variance}
The two-wavelength OPD variance considering multiple Zernike modes is simply the sum over the relevant $i$ in \eq{eq:Z21}.  If that sum includes all $i$ except $i=0$, we obtain the two-wavelength piston-removed OPD variance $\Langle \Delta \ell_{\text{PR}}^2 \Rangle$, which is the total, residual wavefront error that a two-wavelength AO system experiences.  

Unfortunately, it is not possible to derive $\Langle \Delta \ell_{\text{PR}}^2 \Rangle$ in this manner.  An expression for $\Langle \Delta \ell_{\text{PR}}^2 \Rangle$, nevertheless, can be found by evaluating~\cite{Hogge:82,Winocur:83,10529268}
\begin{equation}\label{eq:Z25}
\begin{gathered}
\Langle \Delta \ell_{\text{PR}}^2 \Rangle = \frac{1}{A}\iint_{-\infty}^{\infty} \operatorname{circ}\left(\frac{\rho}{D}\right) \left\langle \left\{ \ell\left(\boldsymbol{\rho},\lambda_\text{B}\right) - a_0\left(\lambda_\text{B}\right) \right. \right. \hfill \\
\left.\left. \quad -\, \left[\ell\left(\boldsymbol{\rho},\lambda_\text{T}\right)-a_0\left(\lambda_\text{T}\right)\right]\right\}^2\right\rangle \text{d}^2\rho, \hfill    
\end{gathered}
\end{equation}
where $a_0$ is the piston Zernike polynomial coefficient.  Note that $Z_0 = 1$ and therefore, does not appear in \eq{eq:Z25}.  

Following the same mathematical steps as in Sections~\ref{sec:thy1}--\ref{sec:thy3}, we obtain a contour integral that is of identical form to \eq{eq:Z21} with $i=0$, except $C$ crosses the real $s$ axis between $-1/12 < \operatorname{Re}\left(s\right) < 0$.  In our chosen definition of a Meijer G-function~\cite{Wolfram}, i.e.,
\begin{widetext}
\begin{equation}\label{eq:G}
\begin{gathered}
G^{m,n}_{p,q}\left(z\left|\begin{matrix} a_1, \cdots, a_n; a_{n+1}, \cdots, a_p\\b_1,\cdots,b_m;b_{m+1},\cdots, b_q\end{matrix}\right.\right) 
 =\frac{1}{\text{j}2\pi} \int_\gamma \frac{\prod_{j=1}^m\Gamma\left(b_j+s\right)\prod_{j=1}^n\Gamma\left(1-a_j-s\right)}{\prod_{j=n+1}^p\Gamma\left(a_j+s\right)\prod_{j=m+1}^q\Gamma\left(1-b_j-s\right)} z^{-s} \text{d}s, 
\end{gathered}
\end{equation}
the contour $\gamma$ passes between the poles of the $\Gamma\left(b_j+s\right)$ and $\Gamma\left(1-a_j-s\right)$ gamma functions.  Here, however, $C$ splits the poles of $\Gamma\left(-5/12+s\right)$ and $\Gamma\left(1/12+s\right)$, leaving the pole at $s = 5/12$ on the right side of $C$.  Therefore, to express $\Langle \Delta \ell_{\text{PR}}^2 \Rangle$ in the form of a Meijer G-function, we need to subtract the $s = 5/12$ pole contribution from $G^{m,n}_{p,q}$.  Performing the requisite complex-plane analysis, we obtain
\begin{equation}\label{eq:Z27}
\begin{gathered}
\Langle \Delta \ell_{\text{PR}}^2 \Rangle = -2^{2/3} \frac{\sqrt{\pi}}{3} \Gamma\begin{bmatrix}5/6,11/6,7/12,17/12\\2/3,11/3\end{bmatrix} C_n^2 z  \sum\limits_{k=2}^5 c_k \alpha_k^{5/6} \hfill \\
\quad -\, \frac{2^{-7/2}}{\sqrt{\pi}} \frac{5}{9} \Gamma\begin{bmatrix}5/6\\2/3,11/3\end{bmatrix} C_n^2 z D^{5/3} \sum\limits_{k=1}^5 c_k  G^{3,5}_{5,7}\left(\frac{D^4}{64 \alpha_k^2}\left|\begin{matrix*}[c] -\dfrac{1}{6}, -\dfrac{2}{3},1,-\dfrac{1}{3},-\dfrac{5}{6}; \text{---}\text{---} \\[10pt] -\dfrac{5}{12}, \dfrac{1}{12},1;-\dfrac{11}{12},-\dfrac{17}{12}, -\dfrac{5}{12},-\dfrac{11}{12}\end{matrix*}\right.\right). 
\end{gathered}
\end{equation}
\end{widetext}

Like above, we obtain the asymptotic approximation for $\Langle \Delta \ell_{\text{PR}}^2 \Rangle$ (in the limit $N_F \to \infty$) by summing contributions from poles to the right of $C$.  Proceeding in ascending order, the pole contribution from $s=0$ is zero.  Conveniently, the dominant contribution comes from the pole at $s = 5/12$.  Expanding the sum over $k$, $\Langle \Delta \ell_{\text{PR}}^2 \Rangle$ is approximately   
\begin{equation}\label{eq:Z28}
\begin{gathered}
\Langle \Delta \ell_{\text{PR}}^2 \Rangle \approx -\frac{\pi^{-1/3}}{6} \Gamma\begin{bmatrix}5/6,11/6,7/12,17/12\\2/3,11/3\end{bmatrix} C_n^2 z^{11/6} \hfill \\
 \;\times \left[2^{-1/6}\left(\lambda_\text{T}^{5/6}+\lambda_\text{B}^{5/6}\right) - \left|\lambda_\text{T}-\lambda_\text{B}\right|^{5/6} - \left(\lambda_\text{T}+\lambda_\text{B}\right)^{5/6}\right]. \hfill
\end{gathered}    
\end{equation}
As we show in Section~\ref{sec:sim}, this is a very good approximation for $\Langle \Delta \ell_{\text{PR}}^2 \Rangle$.  In an earlier work~\cite{10529268}, we derived the plane-wave two-wavelength piston-removed OPD variance.  The spherical- and plane-wave expressions are related by
\begin{equation}
\dfrac{\Langle \Delta \ell_{\text{PR,sw}}^2 \Rangle}{\Langle \Delta \ell_{\text{PR,pw}}^2 \Rangle} = \Gamma\begin{bmatrix}11/6,17/6\\11/3\end{bmatrix}  \approx 0.4043,   
\end{equation}
which is slightly more than in the single-wavelength case, i.e., $3/8$~\cite{Noll:76,Sasiela:07}.

\subsection{Two-wavelength piston- and tilt-removed OPD variance}
Before proceeding to two-wavelength tilt or tracking errors, we can easily derive an expression for the two-wavelength piston- and tilt-removed OPD variance $\Langle \Delta \ell_{\text{PTR}}^2 \Rangle$ using the above analysis.  This quantity is a measure of the total, residual wavefront error due to higher-order aberrations in a two-wavelength AO system.

The two-wavelength piston- and tilt-removed OPD variance is given by
\begin{equation}\label{eq:Z31}
\begin{gathered}
\Langle \Delta \ell_{\text{PTR}}^2 \Rangle = \frac{1}{A}\iint_{-\infty}^{\infty} \operatorname{circ}\left(\frac{\rho}{D}\right) \left\langle \left\{ \ell_\text{PR}\left(\boldsymbol{\rho},\lambda_\text{B}\right) \right. \right. \hfill \\
\left.\left. -\, \sum\limits_{m=1}^2 a_m\left(\lambda_\text{B}\right)Z_m\left(\frac{\rho}{D/2},\phi\right)  - \left[\ell_\text{PR}\left(\boldsymbol{\rho},\lambda_\text{T}\right) \right. \right. \right. \hfill \\
\left. \left.\left. -\, \sum\limits_{m=1}^2 a_m\left(\lambda_\text{T}\right)Z_m\left(\frac{\rho}{D/2},\phi\right) \right]\right\}^2\right\rangle \text{d}^2\rho, \hfill    
\end{gathered}
\end{equation}
where $\ell_\text{PR}\left(\boldsymbol{\rho},\lambda\right) = \ell\left(\boldsymbol{\rho},\lambda\right) - a_0\left(\lambda\right)$ and $Z_1$ and $Z_2$ are the Zernike polynomials for $x$ and $y$ tilts, respectively.  Since Zernike polynomials are orthogonal over circular apertures, \eq{eq:Z31} is equal to
\begin{equation}\label{eq:Z30}
\Langle \Delta \ell_{\text{PTR}}^2 \Rangle = \Langle \Delta \ell_{\text{PR}}^2 \Rangle - 2  \Langle \Delta \ell_1^2 \Rangle,  
\end{equation}
where the factor of two accounts for both axes of tilt.  In \eq{eq:Z30}, either the exact [\eqs{eq:Z27}{eq:Z21}] or asymptotic [\eqs{eq:Z28}{eq:Z23}] relations for $\Langle \Delta \ell_{\text{PR}}^2 \Rangle$ and $\Langle \Delta \ell_1^2 \Rangle$ can be used.

\section{Tilt errors} \label{sec:tlt}
\subsection{Two-wavelength Z-tilt variance}
As it will be useful later on, we quickly derive the two-wavelength Zernike-tilt (Z-tilt) angle variance, namely,
\begin{equation}\label{eq:Ztiltvar}
\Langle \Delta T_Z^2\Rangle = \Langle \left| \boldsymbol{T}_Z\left(\lambda_\text{B}\right) - \boldsymbol{T}_Z\left(\lambda_\text{T}\right) \right|^2 \Rangle,   
\end{equation}
using the above analysis.  We obtain the two-wavelength Z-tilt OPD variance by setting $i = 1$ in either \eq{eq:Z21} or \eqref{eq:Z23} and multiplying the result by two (to include both the $x$ and $y$ tilts).  We can convert this to the more physical tilt angle variance by realizing that the OPD variance (averaged over the circular receiving aperture) is equal to
\begin{equation}
2 \Langle \ell_1^2 \Rangle = \frac{1}{A} \iint_{-\infty}^{\infty} \operatorname{circ}\left(\frac{\rho}{D}\right) \Langle \left(\boldsymbol{T}_Z \cdot \boldsymbol{\rho}\right)^2 \Rangle \text{d}^2 \rho,    
\end{equation}
where $\boldsymbol{T}_Z = \uvecm{x}T_{Zx} + \uvecm{y}T_{Zy}$ is the Z-tilt angle.  The integral is easy to compute and reveals that $2 \Langle \ell_1^2 \Rangle = \left({D}/{4}\right)^2 \Langle T_Z^2 \Rangle$.  Consequently, the two-wavelength Z-tilt angle variance, using the asymptotic result in \eq{eq:Z23}, is
\begin{equation}\label{eq:ztilt}
\begin{gathered}
\Langle \Delta T_Z^2 \Rangle \approx \frac{2^{29/6}}{\pi^{7/3}} \frac{5}{9} \Gamma\begin{bmatrix}5/6,-7/6,13/6,1/6\\11/3,1/4,3/4\end{bmatrix} C_n^2 z^{10/3} D^{-5}  \hfill \\
\; \times \left[2^{4/3}\left(\lambda_\text{T}^{7/3}+\lambda_\text{B}^{7/3}\right)-\left|\lambda_\text{T}-\lambda_\text{B}\right|^{7/3} - \left(\lambda_\text{T}+\lambda_\text{B}\right)^{7/3}\right]. \hfill    
\end{gathered}
\end{equation}

\subsection{Two-wavelength G-tilt variance}
The gradient-tilt (G-tilt) angle is defined as 
\begin{equation}\label{eq:g1}
\boldsymbol{T}_G\left(\lambda\right) = \frac{1}{k A} \iint_{-\infty}^{\infty} \operatorname{circ}\left(\frac{\rho}{D}\right) \nabla \phi\left(\boldsymbol{\rho},\lambda\right) \text{d}^2\rho.
\end{equation}    
Let the difference of this quantity at two wavelengths, $\lambda_\text{B}$ and $\lambda_\text{T}$, be
\begin{equation}\label{eq:g2}
\Delta \boldsymbol{T}_G = \boldsymbol{T}_G\left(\lambda_\text{B}\right) -  \boldsymbol{T}_G\left(\lambda_\text{T}\right).  
\end{equation}
Consequently, the variance of $\Delta \boldsymbol{T}_G$, i.e., the two-wavelength G-tilt angle variance, is
\begin{equation}\label{eq:g3}
\begin{gathered}
\Langle \Delta \boldsymbol{T}_G \cdot \Delta \boldsymbol{T}_G \Rangle = \Langle \left|\boldsymbol{T}_G\left(\lambda_\text{B}\right) -  \boldsymbol{T}_G\left(\lambda_\text{T}\right)\right|^2\Rangle = \Langle \Delta T_G^2 \Rangle  \hfill \\
\quad = \Langle {T}_G^2\left(\lambda_\text{T}\right) \Rangle + \Langle {T}_G^2\left(\lambda_\text{B}\right) \Rangle - 2\Langle\boldsymbol{T}_G\left(\lambda_\text{T}\right) \cdot \boldsymbol{T}_G\left(\lambda_\text{B}\right) \Rangle. \hfill  
\end{gathered}
\end{equation}
Again, we focus on the two-wavelength G-tilt covariance, since both $\Langle {T}_G^2\left(\lambda_\text{T}\right) \Rangle$ and $\Langle {T}_G^2\left(\lambda_\text{B}\right) \Rangle$ can be derived from it.  Using \eq{eq:g1}, we obtain
\begin{equation}\label{eq:g4}
\begin{gathered}
\Langle\boldsymbol{T}_G\left(\lambda_\text{T}\right) \cdot \boldsymbol{T}_G\left(\lambda_\text{B}\right) \Rangle = \frac{1}{k_\text{T} k_\text{B} A^2} \iiiint_{-\infty}^{\infty} \operatorname{circ}\left(\frac{\rho_1}{D}\right) \hfill \\
\quad \times \operatorname{circ}\left(\frac{\rho_2}{D}\right) \Langle \nabla_1 \phi\left(\boldsymbol{\rho}_1,\lambda_\text{T}\right) \cdot  \nabla_2 \phi\left(\boldsymbol{\rho}_2,\lambda_\text{B}\right) \Rangle \text{d}^2 \rho_1 \text{d}^2 \rho_2, \hfill
\end{gathered}
\end{equation}
where $\nabla_{1,2} = \uvecm{x}\partial/\partial x_{1,2} + \uvecm{y}\partial/\partial y_{1,2}$. The moment in \eq{eq:g4} is the two-wavelength phase-gradient covariance function $B_{\nabla \phi}$, which we derive in Appendix~\ref{appA}. 

Making the variable substitutions $\boldsymbol{\rho}' = \boldsymbol{\rho}_1$ and $\boldsymbol{\rho} = \boldsymbol{\rho}_1 - \boldsymbol{\rho}_2$ produces
\begin{equation}\label{eq:g5}
\begin{gathered}
\Langle\boldsymbol{T}_G\left(\lambda_\text{T}\right) \cdot \boldsymbol{T}_G\left(\lambda_\text{B}\right) \Rangle = \frac{1}{k_\text{T} k_\text{B} A} \iint_{-\infty}^{\infty} B_{\nabla \phi}\left(\boldsymbol{\rho},\lambda_\text{B},\lambda_\text{T}\right) \hfill \\
\quad \times \frac{1}{A} \iint_{-\infty}^{\infty} \operatorname{circ}\left(\frac{\rho'}{D}\right) \operatorname{circ}\left(\frac{\left|\boldsymbol{\rho}'-\boldsymbol{\rho}\right|}{D}\right) \text{d}^2\rho' \text{d}^2 \rho, \hfill
\end{gathered}    
\end{equation}
where the integrals over $\boldsymbol{\rho}'$ are equal to the optical transfer function (OTF)
\begin{equation}
\Lambda\left(\frac{\rho}{D}\right) = \frac{2}{\pi}\left[\cos^{-1}\left(\frac{\rho}{D}\right)-\frac{\rho}{D}\sqrt{1-\left(\frac{\rho}{D}\right)^2}\right]\operatorname{circ}\left(\frac{\rho}{2D}\right).
\end{equation}
We obtain the two-wavelength G-tilt covariance by transforming \eq{eq:g5} to polar coordinates:
\begin{equation}\label{eq:g7}
\begin{gathered}
\Langle\boldsymbol{T}_G\left(\lambda_\text{T}\right) \cdot \boldsymbol{T}_G\left(\lambda_\text{B}\right) \Rangle \hfill \\
\quad = \frac{2\pi}{k_\text{T} k_\text{B} A} \int_{0}^{\infty} \rho \Lambda\left(\frac{\rho}{D}\right)  B_{\nabla \phi}\left(\rho,\lambda_\text{B},\lambda_\text{T}\right) \text{d}\rho. \hfill
\end{gathered}
\end{equation}

Substituting \eq{eq:a6} into \eq{eq:g7} and rearranging the integrals produces
\begin{equation}
\begin{gathered}
\Langle\boldsymbol{T}_G\left(\lambda_\text{T}\right) \cdot \boldsymbol{T}_G\left(\lambda_\text{B}\right) \Rangle = 4\pi^2 \int_0^z \int_0^\infty \kappa^3 \left(\frac{\zeta}{z}\right)^2  \Phi_n\left(\kappa,\zeta\right) \hfill \\
\; \times \cos\left[\frac{z}{2k_\text{B}} \frac{\zeta}{z} \left(1-\frac{\zeta}{z}\right)\kappa^2 \right] \cos\left[\frac{z}{2k_\text{T}} \frac{\zeta}{z} \left(1-\frac{\zeta}{z}\right)\kappa^2 \right] \hfill \\
\; \times \frac{2\pi}{A}\int_0^\infty \rho \Lambda\left(\frac{\rho}{D}\right) J_0\left(\frac{\zeta}{z}\kappa\rho\right) \text{d}\rho \text{d}\kappa \text{d}\zeta. \hfill
\end{gathered}    
\end{equation}
The integral over $\rho$ is the Fourier--Bessel transform of the OTF and equals $\operatorname{jinc}^2\left[\kappa D \zeta/\left(2z\right)\right]$, where $\operatorname{jinc}\left(x\right) = 2J_1\left(x\right)/x$.

Like above, assuming Kolmogorov $\Phi_n$, constant $C_n^2$, and after substituting everything back into \eq{eq:g3}, we obtain the following integral expression for the two-wavelength G-tilt angle variance:
\begin{equation}\label{eq:g9}
\begin{gathered}
\Langle \Delta T_G^2 \Rangle = 2^{14/3} \frac{5}{9} \Gamma\begin{bmatrix}5/6\\2/3\end{bmatrix} 
 \sqrt{\pi} C_n^2 D^{-2} \sum\limits_{k=1}^5 c_k \int_{0}^z \int_0^\infty \kappa^{-8/3} \hfill \\
\;\times J_1^2\left(\frac{D}{2} \frac{\zeta}{z}\kappa\right) \cos\left[\alpha_k\frac{\zeta}{z}\left(1-\frac{\zeta}{z}\right)\kappa^2\right] \text{d}\kappa \text{d}\zeta, \hfill      
\end{gathered}
\end{equation}
where $\boldsymbol{c}$ and $\boldsymbol{\alpha}$ are given in \eq{eq:Z17}.  

\subsubsection{Evaluating \eq{eq:g9} using Mellin transforms}
Again, utilizing the Mellin convolution theorem transforms \eq{eq:g9} into a contour integral such that
\begin{equation}\label{eq:g9b}
    \begin{gathered}
        \Langle \Delta T_G^2 \Rangle = \sqrt{\frac{2}{\pi}} \frac{5}{9} \Gamma\begin{bmatrix} 5/6\\2/3,11/3\end{bmatrix} C_n^2 z D^{-1/3} \sum\limits_{k=1}^5 c_k \hfill \\
        \quad \times \frac{1}{\text{j}2\pi}
        \int_C \left(\frac{D^4}{64 \alpha_k^2}\right)^{-s} \Gamma\begin{bmatrix}1/12+s,7/12+s,1+s\end{bmatrix} \hfill \\ \quad
        \times\, \Gamma\begin{bmatrix}2/3-s,7/6-s,-s,4/3-s,11/6-s\\17/12-s,23/12-s,11/12-s,17/12-s\end{bmatrix}\text{d}s \hfill \\        
        \;=\sqrt{\frac{2}{\pi}} \frac{5}{9} \Gamma\begin{bmatrix} 5/6\\2/3,11/3\end{bmatrix} C_n^2 z D^{-1/3} \sum\limits_{k=1}^5 c_k \hfill \\
        \quad \times \,G^{3,5}_{5,7}\left(\frac{D^4}{64 \alpha_k^2}\left|\begin{matrix*}[c] \dfrac{1}{3},-\dfrac{1}{6},1,-\dfrac{1}{3},-\dfrac{5}{6}; \text{---}\text{---} \\[10pt] \dfrac{1}{12},\dfrac{7}{12},1;-\dfrac{5}{12},-\dfrac{11}{12},\dfrac{1}{12},-\dfrac{5}{12}\end{matrix*}\right.\right), \hfill
    \end{gathered}    
\end{equation}
where $C$ crosses the real $s$ axis between $-1/12 < \operatorname{Re}\left(s\right) < 0$.  For brevity, we omitted the details of the integration over $\zeta$ as they are similar to \eq{eq:Z21}.        

\subsubsection{Asymptotic solution}
Like \eq{eq:Z21}, \eq{eq:g9b} converges for all values of its argument when $C$ is closed to the left.  Nevertheless, converges is very slow for large $N_F$.  We can obtain a physical and rapidly converging approximation to \eq{eq:g9b} by including pole contributions to the right of $C$, namely, $s=m,\, m+2/3,\, m+7/6,\, m+4/3$, and $m+11/6$ for $m = 0,\, 1,\, 2, \cdots, M-1$.  The contributions from the poles at $s=0$ and $s=1$ are again zero; the dominate contribution comes from the pole at $s = 2/3$:
\begin{equation}\label{eq:g46}
\begin{gathered}
\Langle \Delta T_G^2 \Rangle \approx \frac{2^{11/6}}{\pi^{4/3}} \frac{5}{9} \Gamma\begin{bmatrix}5/6,5/3,-2/3,7/6\\11/3,1/4,3/4\end{bmatrix} C_n^2 z^{7/3} D^{-3} \hfill \\
\;\times \left[2^{1/3}\left(\lambda_\text{T}^{4/3}+\lambda_\text{B}^{4/3}\right)  - \left|\lambda_\text{T}-\lambda_\text{B}\right|^{4/3} - \left(\lambda_\text{T}+\lambda_\text{B}\right)^{4/3}\right]. \hfill
\end{gathered}    
\end{equation}

\subsection{Two-wavelength G-tilt, Z-tilt variance}
The two-wavelength G-tilt, Z-tilt angle variance is defined as
\begin{equation}\label{eq:gz1}
\begin{gathered}
\Langle \left| \boldsymbol{T}_{G}\left(\lambda_\text{B}\right) - \boldsymbol{T}_{Z}\left(\lambda_\text{T}\right) \right|^2\Rangle = \langle \Delta T_{GZ}^2\rangle  \hfill \\
\quad  = \Langle  {T}^2_{Z}\left(\lambda_\text{T}\right) \Rangle + \Langle  {T}^2_{G}\left(\lambda_\text{B}\right) \Rangle - 2\Langle \boldsymbol{T}_{G}\left(\lambda_\text{B}\right) \cdot \boldsymbol{T}_{Z}\left(\lambda_\text{T}\right) \Rangle, \hfill
\end{gathered}
\end{equation}
where the terms are the Z-tilt and G-tilt angle variances and the G-tilt, Z-tilt covariance.  

We can derive expressions for the Z-tilt and G-tilt angle variances directly from \eqs{eq:Z21}{eq:g9b}, respectively.  For the former,
\begin{equation}\label{eq:gz1a}
\begin{gathered}
\Langle T_Z^2\left(\lambda_\text{T}\right)\Rangle = 2 \left(\frac{4}{D}\right)^2 \Langle \ell_1^2\left(\lambda_\text{T}\right) \Rangle \hfill \\
\; =\frac{2^{3/2}}{\sqrt{\pi}} \frac{5}{9}\Gamma\begin{bmatrix} 5/6\\2/3,11/3\end{bmatrix} C_n^2 z D^{-1/3} \hfill \\
\; \times \sum\limits_{k=1,2} G^{3,5}_{5,7}\left(\frac{D^4}{64 \alpha_k^2}\left|\begin{matrix*}[c] -\dfrac{1}{6}, -\dfrac{2}{3},1,-\dfrac{1}{3},-\dfrac{5}{6}; \text{---}\text{---} \\[10pt] \dfrac{1}{12}, \dfrac{7}{12},1;-\dfrac{17}{12},-\dfrac{23}{12}, -\dfrac{5}{12},-\dfrac{11}{12}\end{matrix*}\right.\right) \hfill
\end{gathered}    
\end{equation}
and the latter,
\begin{equation}\label{eq:gz1b}
\begin{gathered}
\Langle T_G^2\left(\lambda_\text{B}\right)\Rangle = \frac{2^{-1/2}}{\sqrt{{\pi}}} \frac{5}{9} \Gamma\begin{bmatrix} 5/6\\2/3,11/3\end{bmatrix} C_n^2 z D^{-1/3} \hfill \\
\; \times \sum\limits_{k=1,3}G^{3,5}_{5,7}\left(\frac{D^4}{64 \alpha_k^2}\left|\begin{matrix*}[c] \dfrac{1}{3},-\dfrac{1}{6},1,-\dfrac{1}{3},-\dfrac{5}{6}; \text{---}\text{---} \\[10pt] \dfrac{1}{12},\dfrac{7}{12},1;-\dfrac{5}{12},-\dfrac{11}{12},\dfrac{1}{12},-\dfrac{5}{12}\end{matrix*}\right.\right).\hfill
\end{gathered}    
\end{equation}

We now focus on the G-tilt, Z-tilt covariance.  The Z-tilt angle takes the form
\begin{equation}\label{eq:gz2}
\begin{split}
\boldsymbol{T}_{Z}\left(\lambda\right) &= \sum\limits_{j=x,y} \uvecm{j}\frac{4}{kD}a_j\left(\lambda\right) \\
a_j\left(\lambda\right) &= \frac{1}{A} \iint_{-\infty}^{\infty} \operatorname{circ}\left(\frac{\rho}{D}\right) \phi\left(\boldsymbol{\rho},\lambda\right) Z_{1,1}^j\left(\frac{\rho}{D/2},\phi\right) \text{d}^2\rho, 
\end{split}    
\end{equation}
where $Z_{1,1}^j$ is the Zernike polynomial for $j = x,y$ tilt.  Using this definition and that of the G-tilt angle given in \eq{eq:g1}, the G-tilt, Z-tilt covariance equals
\begin{equation}\label{eq:gz3}
\begin{gathered}
\Langle \boldsymbol{T}_{G}\left(\lambda_\text{B}\right) \cdot \boldsymbol{T}_{Z}\left(\lambda_\text{T}\right) \Rangle \hfill \\
\; = \frac{\pi D}{k_\text{B} k_\text{T} A^3} \sum_{j=x,y} \iiiint_{-\infty}^{\infty} \operatorname{circ} \left(\frac{\rho_1}{D}\right) \operatorname{circ} \left(\frac{\rho_2}{D}\right)\hfill \\
\;  \times\, Z_{1,1}^j\left(\frac{\rho_1}{D/2},\phi_1\right) \Langle \phi \left(\boldsymbol{\rho}_1,\lambda_\text{T}\right) \frac{\partial}{\partial j} \phi \left(\boldsymbol{\rho}_2,\lambda_\text{B}\right) \Rangle \text{d}^2\rho_1 \text{d}^2\rho_2. \hfill
\end{gathered}
\end{equation}
The moment in \eq{eq:gz3} is equal to the gradient of $B_S$ and is derived in Appendix~\ref{appB}. 

Making the variable substitutions $\boldsymbol{\rho}' = \boldsymbol{\rho}_1$ and $\boldsymbol{\rho} = \boldsymbol{\rho}_1 - \boldsymbol{\rho}_2$ produces
\begin{equation}\label{eq:gz4}
\begin{gathered}
\Langle \boldsymbol{T}_\text{G}\left(\lambda_\text{B}\right) \cdot \boldsymbol{T}_{Z}\left(\lambda_\text{T}\right) \Rangle \hfill \\
\;= \frac{\pi D}{k_\text{B} k_\text{T} A^2} \sum_{j=x,y} \iint_{-\infty}^{\infty} \left[-\uvecm{j} \cdot \nabla_{\boldsymbol{\rho}} B_S\left(\boldsymbol{\rho},\lambda_\text{B},\lambda_\text{T}\right) \right] \hfill \\
\; \times\frac{1}{A} \iint_{-\infty}^{\infty}  Z_{1,1}^j\left(\frac{\rho'}{D/2},\phi'\right) \operatorname{circ} \left(\frac{\rho'}{D}\right) \hfill \\
\;\times\operatorname{circ} \left(\frac{\left|\boldsymbol{\rho}'-\boldsymbol{\rho}\right|}{D}\right) \text{d}^2\rho' \text{d}^2\rho. \hfill    
\end{gathered}    
\end{equation}
Via the Fourier transform of $Z_{1,1}^j$~\cite{Noll:76,Sasiela:07,Lakshminarayanan10042011}, the integrals over $\boldsymbol{\rho}'$ can be evaluated in closed form such that
\begin{equation}\label{eq:gz5}
\begin{gathered}
\Langle \boldsymbol{T}_\text{G}\left(\lambda_\text{B}\right) \cdot \boldsymbol{T}_{Z}\left(\lambda_\text{T}\right) \Rangle \hfill \\
\;= \frac{\pi D}{k_\text{B} k_\text{T} A} \left(\frac{4}{D}\right)^2 \sum_{j=x,y} \int_0^\infty \kappa^{-1} J_1\left(\frac{D}{2}\kappa\right) J_2\left(\frac{D}{2}\kappa\right) \hfill \\
\;\times\iint_{-\infty}^{\infty} \left(\uvecm{j}\cdot\uvecm{\rho}\right) \left[-\uvecm{j} \cdot \nabla_{\boldsymbol{\rho}} B_S\left(\boldsymbol{\rho},\lambda_\text{B},\lambda_\text{T}\right) \right] J_1\left(\kappa \rho\right) \text{d}^2 \rho \text{d}\kappa. \hfill
\end{gathered}    
\end{equation}

Substituting \eq{eq:b3} into \eq{eq:gz5} and rearranging the integrals reveals
\begin{equation}\label{eq:gz6}
\begin{gathered}
\Langle \boldsymbol{T}_\text{G}\left(\lambda_\text{B}\right) \cdot \boldsymbol{T}_{Z}\left(\lambda_\text{T}\right) \Rangle \hfill \\
\;= \frac{4\pi^2 D}{A} \left(\frac{4}{D}\right)^2 \int_0^z \int_0^\infty \kappa^2 \left(\frac{\zeta}{z}\right) \Phi_n\left(\kappa,\zeta\right) \hfill \\
\; \times\cos\left[\frac{z}{2k_\text{B}} \frac{\zeta}{z} \left(1-\frac{\zeta}{z}\right)\kappa^2 \right]  \cos\left[\frac{z}{2k_\text{T}} \frac{\zeta}{z} \left(1-\frac{\zeta}{z}\right)\kappa^2 \right] \hfill \\
\; \times\left[\sum\limits_{j=x,y} \int_0^{2\pi} \left(\uvecm{j}\cdot\uvecm{\rho}\right)^2 \text{d}\phi\right] \int_0^\infty \frac{1}{\kappa'} J_1\left(\frac{D}{2}\kappa'\right) J_2\left(\frac{D}{2}\kappa'\right)\hfill \\
\; \times\int_0^\infty \rho J_1\left(\kappa' \rho\right) J_1 \left(\frac{\zeta}{z}\kappa \rho\right) \text{d}\rho \text{d}\kappa' \text{d} \kappa \text{d}\zeta. \hfill
\end{gathered}    
\end{equation}
The quantity in brackets equals $2\pi$, and the integral over $\rho$ is the Bessel function orthogonality relation simplifying \eq{eq:gz6} to
\begin{equation}\label{eq:gz7}
\begin{gathered}
\Langle \boldsymbol{T}_\text{G}\left(\lambda_\text{B}\right) \cdot \boldsymbol{T}_{Z}\left(\lambda_\text{T}\right) \Rangle = \frac{32\pi^2}{D}\left(\frac{4}{D}\right)^2 \int_0^z \int_0^\infty \left(\frac{\zeta}{z}\right)^{-1} \hfill \\
\;\times \Phi_n\left(\kappa,\zeta\right)  J_1\left(\frac{D}{2}\frac{\zeta}{z}\kappa\right) J_2\left(\frac{D}{2}\frac{\zeta}{z}\kappa\right) \hfill \\
\;\times \cos\left[\frac{z}{2k_\text{B}} \frac{\zeta}{z} \left(1-\frac{\zeta}{z}\right)\kappa^2 \right]  \cos\left[\frac{z}{2k_\text{T}} \frac{\zeta}{z} \left(1-\frac{\zeta}{z}\right)\kappa^2 \right] \text{d}\kappa \text{d}\zeta. \hfill
\end{gathered}
\end{equation}
Assuming Kolmogorov $\Phi_n$ and constant $C_n^2$, we finally arrive at the two-wavelength G-tilt, Z-tilt covariance: 
\begin{equation}\label{eq:gz8}
\begin{gathered}
\Langle \boldsymbol{T}_\text{G}\left(\lambda_\text{B}\right) \cdot \boldsymbol{T}_{Z}\left(\lambda_\text{T}\right) \Rangle = 2^{20/3} \frac{5}{9} \Gamma\begin{bmatrix}5/6\\2/3\end{bmatrix} 
 \sqrt{\pi} C_n^2 D^{-3} \hfill \\
 \; \times\sum\limits_{k=4,5} \int_{0}^z \int_0^\infty \kappa^{-11/3}\left(\frac{\zeta}{z}\right)^{-1} J_1\left(\frac{D}{2}\frac{\zeta}{z}\kappa\right)  \hfill \\
 \;\times J_2\left(\frac{D}{2}\frac{\zeta}{z}\kappa\right) \cos\left[\alpha_k\frac{\zeta}{z}\left(1-\frac{\zeta}{z}\right)\kappa^2\right] \text{d}\kappa \text{d}\zeta. \hfill      
\end{gathered}    
\end{equation}

\subsubsection{Evaluating \eq{eq:gz8} using Mellin transforms}
Utilizing the Mellin convolution theorem transforms \eq{eq:gz8} into a contour integral such that
\begin{equation}\label{eq:gz9}
    \begin{gathered}
        \Langle \boldsymbol{T}_\text{G}\left(\lambda_\text{B}\right) \cdot \boldsymbol{T}_{Z}\left(\lambda_\text{T}\right) \Rangle = \frac{2^{1/2}}{\sqrt{\pi}} \frac{5}{9} \Gamma\begin{bmatrix} 5/6\\2/3,11/3\end{bmatrix} C_n^2 z D^{-1/3} \hfill \\
        \quad \times\sum\limits_{k=4,5}  \frac{1}{\text{j}2\pi}
        \int_C \left(\frac{D^4}{64 \alpha_k^2}\right)^{-s} \Gamma\begin{bmatrix}1/12+s,7/12+s,1+s\end{bmatrix} \hfill \\ \quad
        \times\, \Gamma\begin{bmatrix}7/6-s,5/3-s,-s,4/3-s,11/6-s\\23/12-s,29/12-s,17/12-s,23/12-s\end{bmatrix}\text{d}s \hfill \\        
        \;=\frac{2^{1/2}}{\sqrt{\pi}} \frac{5}{9} \Gamma\begin{bmatrix} 5/6\\2/3,11/3\end{bmatrix} C_n^2 z D^{-1/3}   \hfill \\
        \quad \times \sum\limits_{k=4,5} G^{3,5}_{5,7}\left(\frac{D^4}{64 \alpha_k^2}\left|\begin{matrix*}[c] -\dfrac{1}{6},-\dfrac{2}{3},1,-\dfrac{1}{3},-\dfrac{5}{6}; \text{---}\text{---} \\[10pt] \dfrac{1}{12},\dfrac{7}{12},1;-\dfrac{11}{12},-\dfrac{17}{12},-\dfrac{5}{12},-\dfrac{11}{12}\end{matrix*}\right.\right), \hfill
    \end{gathered}    
\end{equation}
where $C$ crosses the real $s$ axis between $-1/12 < \operatorname{Re}\left(s\right) < 0$.  To obtain the G-tilt, Z-tilt angle variance, Eqs.~\eqref{eq:gz1a},~\eqref{eq:gz1b}, and~\eqref{eq:gz9} must be substituted back into \eq{eq:gz1}.  

\subsubsection{Asymptotic solution}
As before, we can obtain a physical expression for the two-wavelength G-tilt, Z-tilt angle variance by including contributions from the poles to the right of the integration contours in Eqs.~\eqref{eq:Z21},~\eqref{eq:g9b}, and~\eqref{eq:gz9}, respectively.  The dominant contribution comes from the pole at $s=0$.  Applying Cauchy's integral formula, we obtain
\begin{equation}\label{eq:gz10}
\begin{gathered}
\langle \Delta T_{GZ}^2\rangle \approx \frac{2^{-7/2}}{\sqrt{\pi}} \frac{5}{9} \frac{77}{207}  \Gamma\begin{bmatrix}5/6,1/12,7/12,7/6\\11/3,17/12,23/12\end{bmatrix} \hfill \\
\quad \times\, \Gamma\begin{bmatrix}4/3,11/6\\23/12,29/12\end{bmatrix} C_n^2 z D^{-1/3}. \hfill    
\end{gathered}
\end{equation}
Although \eq{eq:gz10} does not depend on $\lambda_\text{T}$ nor $\lambda_\text{B}$ (the full asymptotic series is presented in Appendix~\ref{appD}), it is a good approximation for the G-tilt, Z-tilt angle variance for physically relevant values of the argument $D^4/\left(64 \alpha_k^2\right)$.

\section{Validation}\label{sec:sim}
\subsection{Simulation setup and procedure}
To validate the above theory, we conducted wave-optics simulations, where we numerically propagated the field emanating from a point source through uniformly distributed atmospheric turbulence to a receiver plane.  We modeled the atmospheric turbulence environment using 20 Kolmogorov phase screens, which were $1 \text{ m} \times 1 \text{ m}$ in size and consisted of $1024 \times {1024}$ points. We generated the Kolmogorov phase screens using the well-known Fourier/spectral method described in Refs.~\cite{doi:10.1088/0959-7174/2/3/003,Frehlich:00,Schmidt:10}. Once in the receiver plane, we collimated the complex-optical field over a circular pupil of diameter $D=30 \text{ cm}$.  We then used that field to calculate $\ell_\text{PR}$, $\ell_\text{PTR}$, $\boldsymbol{T}_Z$, and $\boldsymbol{T}_G$.

For these simulations, we let $z=5$ $\text{km}$, $C_n^2=7.465\times{10}^{-16}$ $\text{m}^{-2/3}$, $\lambda_\text{T}=2$ $\mu\text{m}$, and we varied $\lambda_\text{B}$ from $1$ to $10$ $\mu\text{m}$. To accomplish simulating a wide range of $\lambda_\text{B}$, we first generated the Kolmogorov phase screens at $\lambda_\text{T}$. Then, applying \eq{eq:WF_phase_relation}, we converted the phase screens $\phi_\text{T}$ to OPL screens by $\ell = \phi_\text{T}/k_\text{T}$. Subsequently, we converted the OPL screens back to phase screens for a given $\lambda_\text{B}$ as $\phi_\text{B}=k_\text{B}\ell$. 

We performed 1,000 independent realizations or trials to arrive at statistically meaningful results. The $\ell_\text{PR}$, $\ell_\text{PTR}$, $\boldsymbol{T}_Z$, and $\boldsymbol{T}_G$ we obtained in each trial were used to compute $\Langle \Delta \ell_\text{PR}^2 \Rangle$, $\Langle \Delta \ell_\text{PTR}^2 \Rangle$, $\Langle \Delta T_Z^2 \Rangle$, $\Langle \Delta T_G^2 \Rangle$, and $\Langle \Delta T_{GZ}^2 \Rangle$ using Eqs.~\eqref{eq:Z25}, \eqref{eq:Z31}, \eqref{eq:Ztiltvar}, \eqref{eq:g3}, and \eqref{eq:gz1}, respectively. 

\subsection{Results and discussion}
\begin{figure}
\centering   
\includegraphics*[width=\columnwidth]{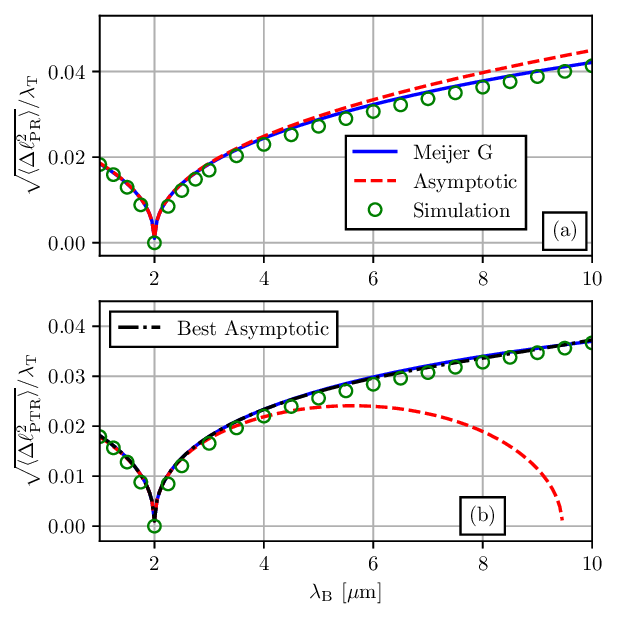}
\caption{Comparison of theoretical and simulated (a) two-wavelength piston-removed $\Langle \Delta \ell_{\text{PR}}^2 \Rangle$ and (b) piston- and tilt-removed $\Langle \Delta \ell_{\text{PTR}}^2 \Rangle$ OPD errors normalized by $\lambda_\text{T}$.}   
\label{fig:Simulation_results1}
\end{figure}

\begin{figure}
\centering   
\includegraphics*[width=\columnwidth]{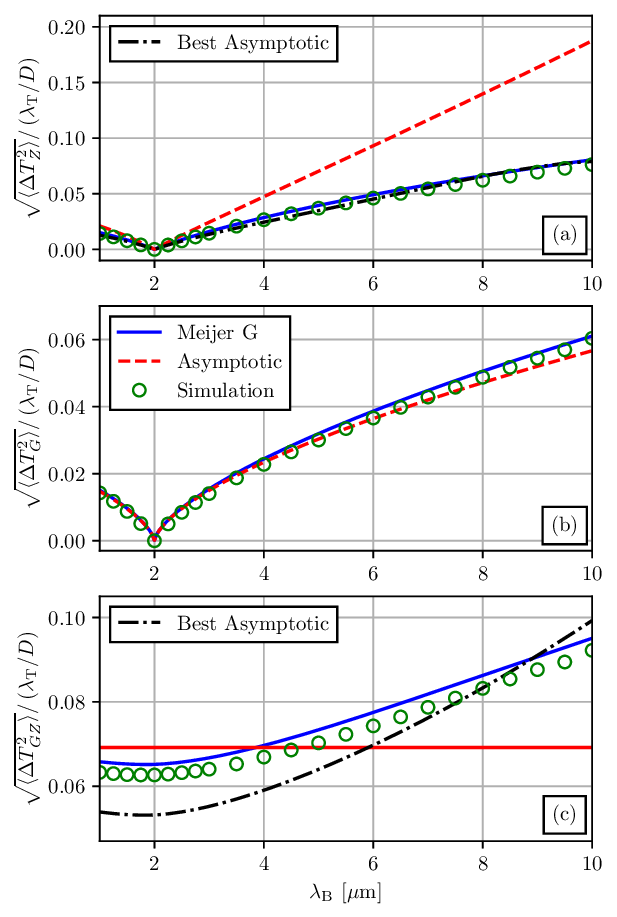}
\caption{Comparison of theoretical and simulated two-wavelength tilt angle errors normalized by the diffraction-limited angular beam width at $\lambda_\text{T}$: (a) Z-tilt, (b) G-tilt, and (c) G-tilt, Z-tilt.}   
\label{fig:Simulation_results2}
\end{figure}

Figures~\ref{fig:Simulation_results1} and~\ref{fig:Simulation_results2} display the simulation results.  In \fig{fig:Simulation_results1}, we present the two-wavelength piston-removed and piston- and tilt-removed OPD errors, normalized by $\lambda_\text{T}$, in (a) and (b), respectively.  Figure~\ref{fig:Simulation_results2} shows the two-wavelength tilt angle errors normalized by the diffraction-limited angular beam width at $\lambda_\text{T}$ in (a)--(c).  In each plot, the blue traces represent the Meijer G-function results, the red-dashed traces are the asymptotic results, and finally, the green circles are the simulation results.  

We also included the ``best asymptotic'' results (black-dashed-dotted traces) in Figs.~\ref{fig:Simulation_results1}(b), \ref{fig:Simulation_results2}(a), and~\ref{fig:Simulation_results2}(c).  The values for $M$, using \eq{eq:c1}, in Figs.~\ref{fig:Simulation_results1}(b) and~\ref{fig:Simulation_results2}(a) were $\left[M_1,\,M_2,\,\cdots,\,M_5\right]  = \left[0,\,1,\,0,\,2,\,1\right]$ and $\left[4,\,2,\,2,\,2,\,3\right]$, respectively.  Likewise, in \fig{fig:Simulation_results2}(c), $\left[M_1,\,M_2,\,M_3\right] = \left[3,2,5\right]$ using Eqs.~\eqref{eq:d1}, \eqref{eq:d3}, and \eqref{eq:d5}.  In all cases, we found the $M$ by brute force: We computed $\Langle \Delta \ell_\text{PTR}^2\Rangle$, $\Langle \Delta T_{Z}^2\Rangle$, and $\Langle \Delta T_{GZ}^2\Rangle$ for every combination of the five and three $M$ for $M\in\left[0,10\right]$ and selected the values that minimized the mean square difference between the asymptotic and Meijer G-function variances.  

The agreement between the simulation results and the exact Meijer G-function expressions is excellent.  The small differences between the two are likely caused by outer scale effects.  The finite grid sizes inherent in the wave-optics simulations manifest as a finite outer scale of turbulence, which truncates the powers in low-spatial-frequency aberrations such as tilt and defocus.  Recall that in the theory, we used the Kolmogorov power spectrum, which has an infinite outer scale.  Consequently, we should physically expect our theory to slightly overpredict the wavefront errors compared to the simulation results.  This is precisely what we observe in \figs{fig:Simulation_results1}{fig:Simulation_results2}.

The accuracies of the $\Langle \Delta \ell_\text{PR}^2 \Rangle$ and $\Langle \Delta T_G^2 \Rangle$ asymptotic expressions [see \eqs{eq:Z28}{eq:g46}] are quite good.  The others rely on the Zernike-mode or Z-tilt asymptotic approximations [Eqs.~\eqref{eq:Z23},~\eqref{eq:ztilt}, or~\eqref{eq:gz10}], which require large values of the argument $D^4/\left(64 \alpha_k^2\right)$ to be accurate.  Indeed, we observe this behavior in Figs.~\ref{fig:Simulation_results1}(b), \ref{fig:Simulation_results2}(a), and~\ref{fig:Simulation_results2}(c), where all results agree around $\lambda_\text{B} \approx \lambda_\text{T}$ and $D^4/\left(64 \alpha_k^2\right)$ is large.  As the $\lambda_\text{B}$-$\lambda_\text{T}$ separation grows and $D^4/\left(64 \alpha_k^2\right)$ becomes smaller, the asymptotic approximations diverge from the others.  Nevertheless, by including more terms in the asymptotic series, we obtain better approximations.  The cost, of course, is the loss of the simple, physical expressions presented in the main text.            

\section{Conclusion}
In this paper, we undertook a rigorous analysis of the relevant wavefront errors necessary to assess two-wavelength AO system performance.  Starting in Section~\ref{sec:thy} with higher-order wavefront errors, we derived expressions for the Zernike-mode, piston-removed, and piston- and tilt-removed variances.  We then proceeded to tilt or tracking errors in Section~\ref{sec:tlt} and derived the Zernike-tilt and gradient-tilt angle variances.  Furthermore, since most AO tracking systems estimate gradient tilt, we derived the gradient-tilt, Zernike-tilt angle variance to also quantify that sensing error.  In both Sections~\ref{sec:thy} and~\ref{sec:tlt}, we used Mellin transform techniques and complex-plane analysis to derive exact expressions for the aforementioned variances in the form of Meijer G-functions.  In addition, we evaluated the resulting contour integrals asymptotically to derive simpler (and more physical) relations for these wavefront errors.  

In Section~\ref{sec:sim}, we performed two-wavelength wave-optics simulations to validate the analysis of the prior sections. We compared predictions made using our theoretical expressions to the simulated results and found them to be in excellent agreement.  The work presented in this paper will be useful in the design and characterization of two-wavelength AO systems.

\appendices
\section{Asymptotic series for the \\Zernike-mode OPD variance}\label{appC}
As described in Section~\ref{sec:asymZ}, an asymptotic approximation (in the limit $N_F \to \infty$) to the two-wavelength Zernike-mode OPD variance [see \eq{eq:Z21}] can be derived by combining the pole contributions that are to the right of $C$ with the steepest descent contribution~\cite{Sasiela:07}.  Further analysis reveals that the latter is negligible compared to the former, and therefore, we obtain the desired result by summing the residues of the poles at $s=m,\, m+7/6,\, m+4/3,\, m+5/3$, and $m+11/6$ for $m = 0,\, 1,\, 2, \cdots, M-1$.  Applying Cauchy's integral formula~\cite{Gbur_2011,Arfken2013} reveals the following:
\begin{equation}\label{eq:c1}
\begin{gathered}
    \Langle \Delta \ell_i^2\Rangle  \approx \frac{2^{-7/2}}{\sqrt{\pi}} \frac{5}{9}\Gamma\begin{bmatrix} 5/6\\2/3,11/3\end{bmatrix}C_n^2 z \left(i+1\right) D^{5/3} \hfill \\ \; \times\sum\limits_{k=2}^5 c_k 
    \left[ S_1\left(m\right) + S_2\left(m+\frac{7}{6}\right) + S_3\left(m+\frac{4}{3}\right) \right. \hfill \\
    \;\left. +\, S_4\left(m+\frac{5}{3}\right)+S_5\left(m+\frac{11}{6}\right)\right].  \hfill
    \end{gathered}
\end{equation}
The sums $S_1$--$S_5$ are
\begin{equation}
\begin{gathered}
S_1\left(m\right) = \sum\limits_{m=2}^{M_1-1} \left(-\frac{64 \alpha_k^2}{D^4}\right)^m  \Gamma\begin{bmatrix} m+i/2-5/12\\-m+i/2+23/12\end{bmatrix}  \hfill \\
\; \times\,\Gamma\begin{bmatrix} m+i/2+1/12,-m+7/6,-m+5/3\\-m+i/2+29/12,-m+17/12\end{bmatrix} \hfill \\
\; \times\, \Gamma\begin{bmatrix}-m+4/3,-m+11/6\\-m+23/12\end{bmatrix}, \hfill
\end{gathered}
\end{equation}
\begin{equation}
\begin{gathered}
S_2\left(m+\frac{7}{6}\right) = \left(\frac{8\alpha_k}{D^2}\right)^{7/3}\sum\limits_{m=0}^{M_2-1} \left(-\frac{64 \alpha_k^2}{D^4}\right)^m  \hfill \\
\; \times\, \Gamma\begin{bmatrix} m+i/2+3/4,m+i/2+5/4,m+5/3\\-m+i/2+3/4,-m+i/2+5/4,-m+1/4\end{bmatrix}  \hfill \\
\; \times\,\Gamma\begin{bmatrix} -m+1/2,-m+1/6,-m+2/3,-m-7/6\\-m+3/4,m+1\end{bmatrix} , \hfill
\end{gathered}
\end{equation}
\begin{equation}
\begin{gathered}
S_3\left(m+\frac{4}{3}\right) = \left(\frac{8\alpha_k}{D^2}\right)^{8/3}\sum\limits_{m=0}^{M_3-1} \left(-\frac{64 \alpha_k^2}{D^4}\right)^m  \hfill \\
\; \times\, \Gamma\begin{bmatrix} m+i/2+11/12,m+i/2+17/12,m+7/3\\-m+i/2+7/12,-m+i/2+13/12,-m+1/12\end{bmatrix}  \hfill \\
\; \times\,\Gamma\begin{bmatrix} -m-1/6,-m+1/3,-m+1/2,-m-4/3\\-m+7/12,m+1\end{bmatrix} , \hfill
\end{gathered}
\end{equation}
\begin{equation}
\begin{gathered}
S_4\left(m+\frac{5}{3}\right) = \left(\frac{8\alpha_k}{D^2}\right)^{10/3}\sum\limits_{m=0}^{M_4-1} \left(-\frac{64 \alpha_k^2}{D^4}\right)^m  \hfill \\
\; \times\, \Gamma\begin{bmatrix} m+i/2+5/4,m+i/2+7/4,m+8/3\\-m+i/2+1/4,-m+i/2+3/4,-m-1/4\end{bmatrix}  \hfill \\
\; \times\,\Gamma\begin{bmatrix} -m-1/2,-m-1/3,-m+1/6,-m-5/3\\-m+1/4,m+1\end{bmatrix} , \hfill
\end{gathered}
\end{equation}
\begin{equation}
\begin{gathered}
S_5\left(m+\frac{11}{6}\right) = \left(\frac{8\alpha_k}{D^2}\right)^{11/3}\sum\limits_{m=0}^{M_5-1} \left(-\frac{64 \alpha_k^2}{D^4}\right)^m  \hfill \\
\; \times\, \Gamma\begin{bmatrix} m+i/2+17/12,m+i/2+23/12,m+17/6\\-m+i/2+1/12,-m+i/2+7/12,-m-5/12\end{bmatrix}  \hfill \\
\; \times\,\Gamma\begin{bmatrix} -m-2/3,-m-1/6,-m-1/2,-m-11/6\\-m+1/12,m+1\end{bmatrix} . \hfill
\end{gathered}
\end{equation}
The values for $M_1$--$M_5$ are easily found through trial and error.  Equation~\eqref{eq:Z23} is equal to \eq{eq:c1} with $M_2=1$ and all others equal to zero.

\section{Two-wavelength phase-gradient \\covariance function}\label{appA}
Let $\phi$ and $\tilde{\phi}$ be the optical phase function and its Fourier transform such that
\begin{equation}\label{eq:a1}
\begin{split}
\tilde{\phi}\left(\boldsymbol{\kappa},\lambda\right) &= \iint_{-\infty}^{\infty} \phi\left(\boldsymbol{\rho},\lambda\right) \exp\left(-\text{j} \boldsymbol{\kappa}\cdot \boldsymbol{\rho}\right) \text{d}^2\rho \\    
{\phi}\left(\boldsymbol{\rho},\lambda\right) &= \frac{1}{\left(2\pi\right)^2}\iint_{-\infty}^{\infty} \tilde{\phi}\left(\boldsymbol{\kappa},\lambda\right) \exp\left(\text{j} \boldsymbol{\kappa}\cdot \boldsymbol{\rho}\right) \text{d}^2\kappa . \\    
\end{split}
\end{equation}
From the second relation in \eq{eq:a1}, the two-wavelength phase-gradient covariance is
\begin{equation}\label{eq:a2}
\begin{gathered}
\Langle \nabla_1 \phi\left(\boldsymbol{\rho}_1,\lambda_\text{B}\right) \cdot  \nabla_2 \phi\left(\boldsymbol{\rho}_2,\lambda_\text{T}\right) \Rangle \hfill \\
\;= \frac{1}{\left(2\pi\right)^4}   \iiiint_{-\infty}^{\infty} \boldsymbol{\kappa}_1 \cdot \boldsymbol{\kappa}_2 \Langle \tilde{\phi}\left(\boldsymbol{\kappa}_1,\lambda_\text{B}\right) \tilde{\phi}^*\left(\boldsymbol{\kappa}_2,\lambda_\text{T}\right) \Rangle  \hfill \\ \; \times \exp\left[\text{j}\left(\boldsymbol{\kappa}_1\cdot \boldsymbol{\rho}_1 -  \boldsymbol{\kappa}_2\cdot \boldsymbol{\rho}_2 \right)\right] \text{d}^2 \kappa_1 \text{d}^2 \kappa_2. \hfill
\end{gathered}
\end{equation}
The moment in \eq{eq:a2} can be found from the first relation in \eq{eq:a1}, namely,
\begin{equation}\label{eq:a3}
\begin{gathered}
 \Langle \tilde{\phi}\left(\boldsymbol{\kappa}_1,\lambda_\text{B}\right) \tilde{\phi}^*\left(\boldsymbol{\kappa}_2,\lambda_\text{T}\right) \Rangle \hfill \\
 \quad=  \iiiint_{-\infty}^{\infty} \Langle \phi\left(\boldsymbol{\rho}_1,\lambda_\text{B}\right) \phi\left(\boldsymbol{\rho}_2,\lambda_\text{T}\right)\Rangle \hfill \\
 \qquad \times\exp\left[-\text{j}\left(\boldsymbol{\kappa}_1\cdot \boldsymbol{\rho}_1 -  \boldsymbol{\kappa}_2\cdot \boldsymbol{\rho}_2 \right)\right] \text{d}^2 \rho_1 \text{d}^2 \rho_2 \hfill \\
 \quad= \iiiint_{-\infty}^{\infty} B_S\left(\boldsymbol{\rho}_1-\boldsymbol{\rho}_2,\lambda_\text{B},\lambda_\text{T}\right) \hfill \\
 \qquad \times\exp\left[-\text{j}\left(\boldsymbol{\kappa}_1\cdot \boldsymbol{\rho}_1 -  \boldsymbol{\kappa}_2\cdot \boldsymbol{\rho}_2 \right)\right] \text{d}^2 \rho_1 \text{d}^2 \rho_2, \hfill
\end{gathered}
\end{equation}
where $B_S$ is the two-wavelength phase covariance function given in \eq{eq:Z11}.  Making the variable substitutions $\boldsymbol{\rho}' = \boldsymbol{\rho}_1$ and $\boldsymbol{\rho} = \boldsymbol{\rho}_1 - \boldsymbol{\rho}_2$ and evaluating the trivial integrals over $\boldsymbol{\rho}'$ yields
\begin{equation}\label{eq:a4}
\begin{gathered}
 \Langle \tilde{\phi}\left(\boldsymbol{\kappa}_1,\lambda_\text{B}\right) \tilde{\phi}^*\left(\boldsymbol{\kappa}_2,\lambda_\text{T}\right) \Rangle \hfill \\
 \quad = \left(2\pi\right)^2 \Phi_S\left(\boldsymbol{\kappa}_2,\lambda_\text{B},\lambda_\text{T}\right) \delta\left(\boldsymbol{\kappa}_1-\boldsymbol{\kappa}_2\right),  
 \end{gathered}
\end{equation}
where $\Phi_S$ is the two-wavelength phase power spectrum and $\delta\left(x\right)$ is the Dirac delta function.  Substituting \eq{eq:a4} into \eq{eq:a2} and evaluating the trivial integrals over $\boldsymbol{\kappa}_2$ produces
\begin{equation}\label{eq:a5}
\begin{gathered}
\Langle \nabla_1 \phi\left(\boldsymbol{\rho}_1,\lambda_\text{B}\right) \cdot  \nabla_2 \phi\left(\boldsymbol{\rho}_2,\lambda_\text{T}\right) \Rangle  
= B_{\nabla \phi}\left(\boldsymbol{\rho},\lambda_\text{B},\lambda_\text{T}\right)\hfill \\
\;= \frac{1}{\left(2\pi\right)^2} \iint_{-\infty}^{\infty} \kappa^2 \Phi_S\left(\boldsymbol{\kappa},\lambda_\text{B},\lambda_\text{T}\right) \exp\left[\text{j}\boldsymbol{\kappa}\cdot\left(\boldsymbol{\rho}_1-\boldsymbol{\rho}_2\right)\right] \text{d}^2 \kappa \hfill \\
\;= - \nabla_{\boldsymbol{\rho}}^2 \left[ \frac{1}{\left(2\pi\right)^2} \iint_{-\infty}^{\infty} \Phi_S\left(\boldsymbol{\kappa},\lambda_\text{B},\lambda_\text{T}\right) \exp\left(\text{j}\boldsymbol{\kappa}\cdot\boldsymbol{\rho}\right) \text{d}^2 \kappa \right] \hfill \\
\;= - \nabla_{\boldsymbol{\rho}}^2 B_S \left(\boldsymbol{\rho},\lambda_\text{B},\lambda_\text{T}\right) , \hfill  
\end{gathered}
\end{equation}
where $\nabla_{\boldsymbol{\rho}}^2$ is the Laplacian with respect to $\boldsymbol{\rho} = \boldsymbol{\rho}_1-\boldsymbol{\rho}_2$.  Substituting \eq{eq:Z11} into \eq{eq:a5} and employing Bessel function identities~\cite{Abramowitz1964,Gradshteyn2000} yields the final result:
\begin{equation}\label{eq:a6}
\begin{gathered}
B_{\nabla \phi}\left(\rho,\lambda_\text{B},\lambda_\text{T}\right) = 4 \pi^2 k_\text{B} k_\text{T} \int_0^z \int_0^\infty \kappa^3 \left(\frac{\zeta}{z}\right)^2 \hfill \\
\; \times\Phi_n\left(\kappa,\zeta\right)  J_0\left(\frac{\zeta}{z}\kappa \rho\right) \cos\left[\frac{z}{2k_\text{B}} \frac{\zeta}{z} \left(1-\frac{\zeta}{z}\right)\kappa^2 \right] \hfill \\
\; \times\cos\left[\frac{z}{2k_\text{T}} \frac{\zeta}{z} \left(1-\frac{\zeta}{z}\right)\kappa^2 \right] \text{d}\kappa \text{d} \zeta. \hfill
\end{gathered}
\end{equation}

\section{Covariance function in \eq{eq:gz3}}\label{appB}
Using the Fourier transform pair in \eq{eq:a1}, the covariance function in \eq{eq:gz3} takes the form
\begin{equation}\label{eq:b1}
\begin{gathered}
\Langle \phi \left(\boldsymbol{\rho}_1,\lambda_\text{T}\right) \frac{\partial}{\partial j} \phi \left(\boldsymbol{\rho}_2,\lambda_\text{B}\right) \Rangle\hfill \\
\;=\frac{-\text{j}}{\left(2\pi\right)^4} \iiiint_{-\infty}^{\infty} \kappa_{2j} 
\Langle \tilde{\phi}\left(\boldsymbol{\kappa}_1,\lambda_\text{T}\right) \tilde{\phi}^*\left(\boldsymbol{\kappa}_2,\lambda_\text{B}\right) \Rangle \hfill \\
\; \times \exp\left[\text{j}\left(\boldsymbol{\kappa}_1\cdot \boldsymbol{\rho}_1 - \boldsymbol{\kappa}_2\cdot \boldsymbol{\rho}_2\right)\right] \text{d}^2\kappa_1 \text{d}^2\kappa_2. \hfill
\end{gathered}
\end{equation}
Substituting \eq{eq:a4} into \eq{eq:b1} and evaluating the trivial integrals over $\boldsymbol{\kappa}_2$ yields
\begin{equation}\label{eq:b2}
\begin{gathered}
\Langle \phi \left(\boldsymbol{\rho}_1,\lambda_\text{T}\right) \frac{\partial}{\partial j} \phi \left(\boldsymbol{\rho}_2,\lambda_\text{B}\right) \Rangle \hfill \\
\;= \frac{-\text{j}}{\left(2\pi\right)^2} \iint_{-\infty}^{\infty} \kappa_j \Phi_S\left(\boldsymbol{\kappa},\lambda_\text{B},\lambda_\text{T}\right) \exp\left[\text{j}\boldsymbol{\kappa} \cdot\left(\boldsymbol{\rho}_1-\boldsymbol{\rho}_2\right)\right] \text{d}^2\kappa \hfill \\
\;= -\frac{\partial}{\partial j} \left[ \frac{1}{\left(2\pi\right)^2} \iint_{-\infty}^{\infty} \Phi_S\left(\boldsymbol{\kappa},\lambda_\text{B},\lambda_\text{T}\right) \exp\left(\text{j}\boldsymbol{\kappa} \cdot\ \boldsymbol{\rho}\right) \right] \hfill \\
\;= -\uvecm{j} \cdot \nabla_{\boldsymbol{\rho}} B_S\left(\boldsymbol{\rho},\lambda_\text{B},\lambda_\text{T}\right). \hfill
\end{gathered}
\end{equation}
Substituting \eq{eq:Z11} into \eq{eq:b2} and employing Bessel function identities~\cite{Abramowitz1964,Gradshteyn2000} produces the final result:
\begin{equation}\label{eq:b3}
\begin{gathered}
 \Langle \phi \left(\boldsymbol{\rho}_1,\lambda_\text{T}\right) \frac{\partial}{\partial j} \phi \left(\boldsymbol{\rho}_2,\lambda_\text{B}\right) \Rangle =  \left( \uvecm{j} \cdot \uvecm{\rho} \right) 4 \pi^2 k_\text{B} k_\text{T}  \hfill \\
 \;\times\int_0^z \int_0^\infty \kappa^2 \left(\frac{\zeta}{z}\right)  \Phi_n\left(\kappa,\zeta\right)  J_1\left(\frac{\zeta}{z}\kappa \rho\right) \hfill \\
 \;\times \cos\left[\frac{z}{2k_\text{B}} \frac{\zeta}{z} \left(1-\frac{\zeta}{z}\right)\kappa^2 \right]  \cos\left[\frac{z}{2k_\text{T}} \frac{\zeta}{z} \left(1-\frac{\zeta}{z}\right)\kappa^2 \right] \text{d}\kappa \text{d} \zeta. \hfill
\end{gathered}    
\end{equation}

\section{Asymptotic series for the \\G-tilt, Z-tilt angle variance}\label{appD}
We begin with the Z-tilt angle variance
\begin{equation}\label{eq:d1}
\begin{gathered}
\Langle T_Z^2\left(\lambda_\text{T}\right)\Rangle = \frac{40}{9} \sqrt{\pi} \Gamma\begin{bmatrix} 5/6\\2/3,11/3\end{bmatrix} C_n^2 z D^{-1/3} \hfill \\
\;\times \sum\limits_{k=1,2} \frac{1}{\text{j}2\pi} \int_C \left(\frac{D^2}{2\alpha_k}\right)^{-s/2} \Gamma\begin{bmatrix} s/2+1/6,s/2+1\\s/4+1/2\end{bmatrix} \hfill \\
\;  \times\, \Gamma\begin{bmatrix} -s/4,-s/2+7/3,-s/2+8/3\\-s/2+17/6,-s/2+29/6\end{bmatrix}\text{d}s. \hfill 
\end{gathered}   
\end{equation}
Note that by making the substitution $s = 4s'$ and applying the Gauss--Legendre multiplication formula, we obtain an expression similar to \eq{eq:Z21}.  Here, we use this form because it makes deriving the asymptotic series easier.

The contour $C$ crosses the real $s$ axis between $-1/3 < \operatorname{Re}\left(s\right) < 0$.  The integral converges for all values of the argument if closed to the left.  We derive the asymptotic series (for large values of the argument) by summing pole contributions that are to the right of $C$, namely, $s = 4m$, $2m+14/3$, and $2m+16/3$ for $m = 0,\,1,\,\cdots,\, M-1$.  Applying Cauchy's integral formula, we obtain 
\begin{equation}\label{eq:d2}
\begin{gathered}
\Langle T_Z^2\left(\lambda_\text{T}\right)\Rangle \approx \frac{80}{9} \sqrt{\pi} \Gamma\begin{bmatrix} 5/6\\2/3,11/3\end{bmatrix} C_n^2 z D^{-1/3} \hfill \\
\;\times \sum\limits_{k=1,2} \left[S_{Z_1}\left(4m\right) + S_{Z_2}\left(2m+14/3\right) + S_{Z_3}\left(2m+16/3\right)\right]. \hfill
\end{gathered}
\end{equation}
The sums $S_{Z_1}$, $S_{Z_2}$, and $S_{Z_3}$ are
\begin{equation}
\begin{gathered}
 S_{Z_1}\left(4m\right) = 2 \sum_{m=0}^{M_1-1} \left(-\frac{D^4}{4\alpha_k^2}\right)^{-m}  \Gamma\begin{bmatrix} 2m+1/6,2m+1\\m+1/2,m+1\end{bmatrix} \hfill \\
 \; \times \,  \Gamma\begin{bmatrix} -2m+7/3,-2m+8/3\\-2m+17/6,-2m+29/6\end{bmatrix}, \hfill
\end{gathered}    
\end{equation}
\begin{equation}
\begin{gathered}
 S_{Z_2}\left(2m+14/3\right) = \left(\frac{D^2}{2\alpha_k}\right)^{-7/3} \sum_{m=0}^{M_2-1} \left(-\frac{D^2}{2\alpha_k}\right)^{-m}  \hfill \\
 \; \times \, \Gamma\begin{bmatrix} m+5/2,m+10/3\\m/2+5/3,m+1\end{bmatrix}  \Gamma\begin{bmatrix} -m/2-7/6,-m+1/3\\-m+1/2,-m+5/2\end{bmatrix}, \hfill
\end{gathered}    
\end{equation}
\begin{equation}
\begin{gathered}
 S_{Z_3}\left(2m+16/3\right) = \left(\frac{D^2}{2\alpha_k}\right)^{-8/3} \sum_{m=0}^{M_3-1} \left(-\frac{D^2}{2\alpha_k}\right)^{-m}  \hfill \\
 \; \times \, \Gamma\begin{bmatrix} m+17/6,m+11/3\\m/2+11/6,m+1\end{bmatrix}  \Gamma\begin{bmatrix} -m/2-4/3,-m-1/3\\-m+1/6,-m+13/6\end{bmatrix}. \hfill
\end{gathered}    
\end{equation}

Proceeding to the G-tilt angle variance,
\begin{equation}\label{eq:d3}
\begin{gathered}
\Langle T_G^2\left(\lambda_\text{B}\right)\Rangle = \frac{5}{18} \sqrt{\pi} \Gamma\begin{bmatrix} 5/6\\2/3,11/3\end{bmatrix} C_n^2 z D^{-1/3} \hfill \\
\;\times \sum\limits_{k=1,2} \frac{1}{\text{j}2\pi} \int_C \left(\frac{D^2}{2\alpha_k}\right)^{-s/2} \Gamma\begin{bmatrix} s/2+1/6,s/2+1\\s/4+1/2\end{bmatrix} \hfill \\
\;  \times\, \Gamma\begin{bmatrix} -s/4,-s/2+4/3,-s/2+8/3\\-s/2+11/6,-s/2+17/6\end{bmatrix}\text{d}s, \hfill 
\end{gathered}   
\end{equation}
where again $C$ is between $-1/3 < \operatorname{Re}\left(s\right) < 0$.  Like above, we derive the asymptotic series by summing contributions from poles to the right of $C$: $s = 4m$, $2m+8/3$, and $2m+16/3$ for $m = 0,\,1,\,\cdots,\, M-1$.  Performing the necessary complex analysis yields
\begin{equation}\label{eq:d4}
\begin{gathered}
\Langle T_G^2\left(\lambda_\text{B}\right)\Rangle \approx \frac{5}{9} \sqrt{\pi} \Gamma\begin{bmatrix} 5/6\\2/3,11/3\end{bmatrix} C_n^2 z D^{-1/3} \hfill \\
\;\times \sum\limits_{k=1,3} \left[S_{G_1}\left(4m\right) + S_{G_2}\left(2m+8/3\right) + S_{G_3}\left(2m+16/3\right)\right]. \hfill
\end{gathered}
\end{equation}
The sums $S_{G_1}$, $S_{G_2}$, and $S_{G_3}$ are
\begin{equation}
\begin{gathered}
 S_{G_1}\left(4m\right) = 2 \sum_{m=0}^{M_1-1} \left(-\frac{D^4}{4\alpha_k^2}\right)^{-m}  \Gamma\begin{bmatrix} 2m+1/6,2m+1\\m+1/2,m+1\end{bmatrix} \hfill \\
 \; \times \,  \Gamma\begin{bmatrix} -2m+4/3,-2m+8/3\\-2m+11/6,-2m+17/6\end{bmatrix}, \hfill
\end{gathered}    
\end{equation}
\begin{equation}
\begin{gathered}
 S_{G_2}\left(2m+8/3\right) = \left(\frac{D^2}{2\alpha_k}\right)^{-4/3} \sum_{m=0}^{M_2-1} \left(-\frac{D^2}{2\alpha_k}\right)^{-m}  \hfill \\
 \; \times \, \Gamma\begin{bmatrix} m+3/2,m+7/3\\m/2+7/6,m+1\end{bmatrix}  \Gamma\begin{bmatrix} -m/2-2/3,-m+4/3\\-m+1/2,-m+3/2\end{bmatrix}, \hfill
\end{gathered}    
\end{equation}
\begin{equation}
\begin{gathered}
 S_{Z_3}\left(2m+16/3\right) = \left(\frac{D^2}{2\alpha_k}\right)^{-8/3} \sum_{m=0}^{M_3-1} \left(-\frac{D^2}{2\alpha_k}\right)^{-m}  \hfill \\
 \; \times \, \Gamma\begin{bmatrix} m+17/6,m+11/3\\m/2+11/6,m+1\end{bmatrix}  \Gamma\begin{bmatrix} -m/2-4/3,-m-4/3\\-m-5/6,-m+1/6\end{bmatrix}. \hfill
\end{gathered}    
\end{equation}

Finally, the G-tilt, Z-tilt covariance is
\begin{equation}\label{eq:d4a}
\begin{gathered}
  \Langle \boldsymbol{T}_{G}\left(\lambda_\text{B}\right) \cdot \boldsymbol{T}_{Z}\left(\lambda_\text{T}\right) \Rangle =  \frac{10}{9} \sqrt{\pi} \Gamma\begin{bmatrix} 5/6\\2/3,11/3\end{bmatrix} C_n^2 z D^{-1/3} \hfill \\
\;\times \sum\limits_{k=4,5} \frac{1}{\text{j}2\pi} \int_C \left(\frac{D^2}{2\alpha_k}\right)^{-s/2} \Gamma\begin{bmatrix} s/2+1/6,s/2+1\\s/4+1/2\end{bmatrix} \hfill \\
\;  \times\, \Gamma\begin{bmatrix} -s/4,-s/2+7/3,-s/2+8/3\\-s/2+17/6,-s/2+23/6\end{bmatrix}\text{d}s, \hfill  
\end{gathered}    
\end{equation}
where $C$ is the same as \eqs{eq:d1}{eq:d3}.  Again, we sum the contributions from the poles to the right of $C$, i.e., $s = 4m$, $2m+14/3$, and $2m=16/3$ for $m = 0,\,1,\,\cdots,\, M-1$, and obtain
\begin{equation}\label{eq:d5}
\begin{gathered}
\Langle \boldsymbol{T}_{G}\left(\lambda_\text{B}\right) \cdot \boldsymbol{T}_{Z}\left(\lambda_\text{T}\right) \Rangle \approx \frac{20}{9} \sqrt{\pi} \Gamma\begin{bmatrix} 5/6\\2/3,11/3\end{bmatrix} C_n^2 z D^{-1/3} \hfill \\
\;\times \sum\limits_{k=4,5} \left[S_{GZ_1}\left(4m\right) + S_{GZ_2}\left(2m+14/3\right) \right. \hfill \\ 
\left.\quad +\, S_{GZ_3}\left(2m+16/3\right)\right]. \hfill
\end{gathered}
\end{equation}
The sums $S_{GZ_1}$, $S_{GZ_2}$, and $S_{GZ_3}$ are
\begin{equation}
\begin{gathered}
 S_{GZ_1}\left(4m\right) = 2 \sum_{m=0}^{M_1-1} \left(-\frac{D^4}{4\alpha_k^2}\right)^{-m}  \Gamma\begin{bmatrix} 2m+1/6,2m+1\\m+1/2,m+1\end{bmatrix} \hfill \\
 \; \times \,  \Gamma\begin{bmatrix} -2m+7/3,-2m+8/3\\-2m+17/6,-2m+23/6\end{bmatrix}, \hfill
\end{gathered}    
\end{equation}
\begin{equation}
\begin{gathered}
 S_{GZ_2}\left(2m+14/3\right) = \left(\frac{D^2}{2\alpha_k}\right)^{-7/3} \sum_{m=0}^{M_2-1} \left(-\frac{D^2}{2\alpha_k}\right)^{-m}  \hfill \\
 \; \times \, \Gamma\begin{bmatrix} m+5/2,m+10/3\\m/2+5/3,m+1\end{bmatrix}  \Gamma\begin{bmatrix} -m/2-7/6,-m+1/3\\-m+1/2,-m+3/2\end{bmatrix}, \hfill
\end{gathered}    
\end{equation}
\begin{equation}
\begin{gathered}
 S_{GZ_3}\left(2m+16/3\right) = \left(\frac{D^2}{2\alpha_k}\right)^{-8/3} \sum_{m=0}^{M_3-1} \left(-\frac{D^2}{2\alpha_k}\right)^{-m}  \hfill \\
 \; \times \, \Gamma\begin{bmatrix} m+17/6,m+11/3\\m/2+11/6,m+1\end{bmatrix}  \Gamma\begin{bmatrix} -m/2-4/3,-m-1/3\\-m+1/6,-m+7/6\end{bmatrix}. \hfill
\end{gathered}    
\end{equation}

We obtain the G-tilt, Z-tilt angle variance by substituting Eqs.~\eqref{eq:d2}, \eqref{eq:d4}, and \eqref{eq:d5} into \eq{eq:gz1}.  Equation~\eqref{eq:gz10} is the result when $\left[M_1,\,M_2,\,M_3\right] = \left[1,0,0\right]$.  

\section*{Acknowledgment}
M.W.H.: The views expressed in this paper are those of the author and do not reflect the policy or position of Epsilon C5I or Epsilon Systems. 

\section*{Disclosures}
The authors declare no conflicts of interest. The U.S. Government is authorized to reproduce and distribute reprints for governmental purposes notwithstanding any copyright notation thereon. Distribution Statement A. Approved for public release: distribution is unlimited. Public Affairs release approval \#: NSWCDD-PN-25-00101.

\bibliographystyle{IEEEtran}
\bibliography{IEEEabrv,main}

\begin{thebibliography}{10}
\providecommand{\url}[1]{#1}
\csname url@samestyle\endcsname
\providecommand{\newblock}{\relax}
\providecommand{\bibinfo}[2]{#2}
\providecommand{\BIBentrySTDinterwordspacing}{\spaceskip=0pt\relax}
\providecommand{\BIBentryALTinterwordstretchfactor}{4}
\providecommand{\BIBentryALTinterwordspacing}{\spaceskip=\fontdimen2\font plus
\BIBentryALTinterwordstretchfactor\fontdimen3\font minus \fontdimen4\font\relax}
\providecommand{\BIBforeignlanguage}[2]{{%
\expandafter\ifx\csname l@#1\endcsname\relax
\typeout{** WARNING: IEEEtran.bst: No hyphenation pattern has been}%
\typeout{** loaded for the language `#1'. Using the pattern for}%
\typeout{** the default language instead.}%
\else
\language=\csname l@#1\endcsname
\fi
#2}}
\providecommand{\BIBdecl}{\relax}
\BIBdecl

\bibitem{Hardy}
J.~W. Hardy, \emph{Adaptive Optics for Astronomical Telescopes}.\hskip 1em plus 0.5em minus 0.4em\relax New York, New York: Oxford University Press, 1998.

\bibitem{Roddier}
F.~Roddier, Ed., \emph{Adaptive Optics in Astronomy}.\hskip 1em plus 0.5em minus 0.4em\relax New York, New York: Cambridge University Press, 1999.

\bibitem{Perram}
G.~P. Perram, S.~J. Cusumano, R.~L. Hengehold, and S.~T. Fiorino, \emph{Introduction to Laser Weapon Systems}.\hskip 1em plus 0.5em minus 0.4em\relax Albuquerque, New Mexico: Directed Energy Professional Society, 2010.

\bibitem{PAO}
R.~K. Tyson, \emph{Principles of Adaptive Optics}, 4th~ed.\hskip 1em plus 0.5em minus 0.4em\relax Boca Raton, Florida: CRC Press, 2015.

\bibitem{Merritt}
P.~H. Merritt and M.~F. Spencer, \emph{Beam Control for Laser Systems}, 2nd~ed.\hskip 1em plus 0.5em minus 0.4em\relax Albuquerque, New Mexico: Directed Energy Professional Society, 2018.

\bibitem{Fugate:23}
\BIBentryALTinterwordspacing
R.~Q. Fugate, J.~D. Barchers, and B.~L. Ellerbroek, ``{David L. Fried: Bringing} vision to atmospheric optics,'' \emph{Appl. Opt.}, vol.~62, no.~23, pp. G112--G127, Aug 2023. [Online]. Available: \url{https://opg.optica.org/ao/abstract.cfm?URI=ao-62-23-G112}
\BIBentrySTDinterwordspacing

\bibitem{Hogge:82}
\BIBentryALTinterwordspacing
C.~B. Hogge and R.~R. Butts, ``Effects of using different wavelengths in wave-front sensing and correction,'' \emph{J. Opt. Soc. Am.}, vol.~72, no.~5, pp. 606--609, May 1982. [Online]. Available: \url{https://opg.optica.org/abstract.cfm?URI=josa-72-5-606}
\BIBentrySTDinterwordspacing

\bibitem{Sasiela:07}
R.~J. Sasiela, \emph{Electromagnetic Wave Propagation in Turbulence}, 2nd~ed.\hskip 1em plus 0.5em minus 0.4em\relax Bellingham, Washington: SPIE Press, 2007.

\bibitem{10529268}
M.~W. Hyde, M.~Kalensky, and M.~F. Spencer, ``Phase error scaling law in two-wavelength adaptive optics,'' \emph{IEEE Photonics Technol. Lett.}, vol.~36, no.~12, pp. 779--782, June 2024.

\bibitem{Fried:98}
\BIBentryALTinterwordspacing
D.~L. Fried, ``Branch point problem in adaptive optics,'' \emph{J. Opt. Soc. Am. A}, vol.~15, no.~10, pp. 2759--2768, Oct 1998. [Online]. Available: \url{https://opg.optica.org/josaa/abstract.cfm?URI=josaa-15-10-2759}
\BIBentrySTDinterwordspacing

\bibitem{Barchers:2003}
\BIBentryALTinterwordspacing
J.~D. Barchers, D.~L. Fried, D.~J. Link, G.~A. Tyler, W.~Moretti, T.~J. Brennan, and R.~Q. Fugate, ``{Performance of wavefront sensors in strong scintillation},'' in \emph{Adaptive Optical System Technologies II}, P.~L. Wizinowich and D.~Bonaccini, Eds., vol. 4839, International Society for Optics and Photonics.\hskip 1em plus 0.5em minus 0.4em\relax SPIE, 2003, pp. 217--227. [Online]. Available: \url{https://doi.org/10.1117/12.457126}
\BIBentrySTDinterwordspacing

\bibitem{Venema:08_2}
\BIBentryALTinterwordspacing
T.~M. Venema and J.~D. Schmidt, ``Optical phase unwrapping in the presence of branch points.'' \emph{Opt. Express}, vol.~16, no.~10, pp. 6985--6998, May 2008. [Online]. Available: \url{http://www.opticsexpress.org/abstract.cfm?URI=oe-16-10-6985}
\BIBentrySTDinterwordspacing

\bibitem{Steinbock:14}
\BIBentryALTinterwordspacing
M.~J. Steinbock, M.~W. Hyde, and J.~D. Schmidt, ``{LSPV+7}, a branch-point-tolerant reconstructor for strong turbulence adaptive optics,'' \emph{Appl. Opt.}, vol.~53, no.~18, pp. 3821--3831, Jun 2014. [Online]. Available: \url{https://opg.optica.org/ao/abstract.cfm?URI=ao-53-18-3821}
\BIBentrySTDinterwordspacing

\bibitem{Banet:20}
\BIBentryALTinterwordspacing
M.~T. Banet and M.~F. Spencer, ``Compensated-beacon adaptive optics using least-squares phase reconstruction,'' \emph{Opt. Express}, vol.~28, no.~24, pp. 36,902--36,914, Nov 2020. [Online]. Available: \url{https://opg.optica.org/oe/abstract.cfm?URI=oe-28-24-36902}
\BIBentrySTDinterwordspacing

\bibitem{Spencer:21}
\BIBentryALTinterwordspacing
M.~F. Spencer, ``Limitations of the deep-turbulence problem,'' in \emph{OSA Imaging and Applied Optics Congress 2021 (3D, COSI, DH, ISA, pcAOP)}.\hskip 1em plus 0.5em minus 0.4em\relax Optica Publishing Group, 2021, p. PW3F.1. [Online]. Available: \url{https://opg.optica.org/abstract.cfm?URI=pcAOP-2021-PW3F.1}
\BIBentrySTDinterwordspacing

\bibitem{Spencer:22}
\BIBentryALTinterwordspacing
M.~F. Spencer and T.~J. Brennan, ``Deep-turbulence phase compensation using tiled arrays,'' \emph{Opt. Express}, vol.~30, no.~19, pp. 33,739--33,755, Sep 2022. [Online]. Available: \url{https://opg.optica.org/oe/abstract.cfm?URI=oe-30-19-33739}
\BIBentrySTDinterwordspacing

\bibitem{Hyde:2024}
\BIBentryALTinterwordspacing
M.~W. Hyde, J.~E. McCrae, M.~Kalensky, and M.~F. Spencer, ``{`Hidden phase'} in two-wavelength adaptive optics,'' \emph{Appl. Opt.}, vol.~63, no.~16, pp. E1--E9, Jun 2024. [Online]. Available: \url{https://opg.optica.org/ao/abstract.cfm?URI=ao-63-16-E1}
\BIBentrySTDinterwordspacing

\bibitem{Kalensky:24}
\BIBentryALTinterwordspacing
M.~Kalensky, D.~Getts, M.~T. Banet, D.~J. Burrell, M.~W. Hyde, and M.~F. Spencer, ``Limitations of beam-control compensation,'' \emph{Opt. Express}, vol.~32, no.~24, pp. 42,301--42,317, Nov 2024. [Online]. Available: \url{https://opg.optica.org/oe/abstract.cfm?URI=oe-32-24-42301}
\BIBentrySTDinterwordspacing

\bibitem{Tyson:82}
\BIBentryALTinterwordspacing
R.~K. Tyson, ``Using the deformable mirror as a spatial filter: application to circular beams,'' \emph{Appl. Opt.}, vol.~21, no.~5, pp. 787--793, Mar 1982. [Online]. Available: \url{https://opg.optica.org/ao/abstract.cfm?URI=ao-21-5-787}
\BIBentrySTDinterwordspacing

\bibitem{Noll:76}
R.~J. Noll, ``Zernike polynomials and atmospheric turbulence,'' \emph{J. Opt. Soc. Am.}, vol.~66, no.~3, pp. 207--211, Mar 1976.

\bibitem{Tyler:82}
\BIBentryALTinterwordspacing
G.~A. Tyler and D.~L. Fried, ``Image-position error associated with a quadrant detector,'' \emph{J. Opt. Soc. Am.}, vol.~72, no.~6, pp. 804--808, Jun 1982. [Online]. Available: \url{https://opg.optica.org/abstract.cfm?URI=josa-72-6-804}
\BIBentrySTDinterwordspacing

\bibitem{Burrell:23}
\BIBentryALTinterwordspacing
D.~J. Burrell, M.~F. Spencer, M.~K. Beason, and R.~G. Driggers, ``Active-tracking scaling laws using the noise-equivalent angle due to speckle,'' \emph{J. Opt. Soc. Am. A}, vol.~40, no.~5, pp. 904--913, May 2023. [Online]. Available: \url{https://opg.optica.org/josaa/abstract.cfm?URI=josaa-40-5-904}
\BIBentrySTDinterwordspacing

\bibitem{Mitchell:25}
\BIBentryALTinterwordspacing
E.~W. Mitchell, D.~J. Burrell, M.~W. Hyde, R.~G. Driggers, and M.~F. Spencer, ``Scaling laws for the noise-equivalent angle and {C-tilt,} {G-tilt} anisoplanatism due to scintillation,'' \emph{Appl. Opt.}, vol.~64, no.~18, pp. E11--E19, Jun 2025. [Online]. Available: \url{https://opg.optica.org/ao/abstract.cfm?URI=ao-64-18-E11}
\BIBentrySTDinterwordspacing

\bibitem{Yura:85}
\BIBentryALTinterwordspacing
H.~T. Yura and M.~T. Tavis, ``Centroid anisoplanatism,'' \emph{J. Opt. Soc. Am. A}, vol.~2, no.~5, pp. 765--773, May 1985. [Online]. Available: \url{https://opg.optica.org/josaa/abstract.cfm?URI=josaa-2-5-765}
\BIBentrySTDinterwordspacing

\bibitem{OIAWA}
V.~N. Mahajan, \emph{Optical Imaging and Aberrations, Part III: Wavefront Analysis}.\hskip 1em plus 0.5em minus 0.4em\relax Bellingham, Washington: SPIE Press, 2013.

\bibitem{Lakshminarayanan10042011}
V.~Lakshminarayanan and A.~Fleck, ``Zernike polynomials: {A} guide,'' \emph{J. Mod. Opt.}, vol.~58, no.~7, pp. 545--561, 2011.

\bibitem{1140133}
A.~Ishimaru, ``Temporal frequency spectra of multifrequency waves in turbulent atmosphere,'' \emph{IEEE Trans. Antennas Propag.}, vol.~20, no.~1, pp. 10--19, Jan 1972.

\bibitem{Ishimaru:99}
------, \emph{Wave Propagation and Scattering in Random Media}.\hskip 1em plus 0.5em minus 0.4em\relax Piscataway, New Jersey: IEEE Press, 1999.

\bibitem{Andrews:05}
L.~C. Andrews and R.~L. Phillips, \emph{Laser Beam Propagation through Random Media}, 2nd~ed.\hskip 1em plus 0.5em minus 0.4em\relax Bellingham, Washington: SPIE Press, 2005.

\bibitem{Tatarskii:61}
V.~I. Tatarskii, \emph{Wave Propagation in a Turbulent Medium}.\hskip 1em plus 0.5em minus 0.4em\relax New York, New York: McGraw-Hill, 1961.

\bibitem{Abramowitz1964}
M.~Abramowitz and I.~A. Stegun, Eds., \emph{Handbook of Mathematical Functions With Formulas, Graphs, and Mathematical Tables}.\hskip 1em plus 0.5em minus 0.4em\relax Washington, DC: National Bureau of Standards, 1964.

\bibitem{Gradshteyn2000}
I.~S. Gradshteyn and I.~M. Ryzhik, \emph{Table of Integrals, Series, and Products}, 8th~ed.\hskip 1em plus 0.5em minus 0.4em\relax Waltham, Massachusetts: Academic Press, 2015.

\bibitem{Brychkov2018}
Y.~A. Brychkov, O.~I. Marichev, and N.~V. Savischenko, \emph{Handbook of Mellin Transforms}.\hskip 1em plus 0.5em minus 0.4em\relax Boca Raton, Florida: CRC Press, 2018.

\bibitem{Wolfram}
\BIBentryALTinterwordspacing
{Wolfram Research, Inc.}, ``{Meijer G-Function}.'' [Online]. Available: \url{https://mathworld.wolfram.com/MeijerG-Function.html}
\BIBentrySTDinterwordspacing

\bibitem{Luke1975}
Y.~L. Luke, \emph{Mathematical Functions and Their Approximations}.\hskip 1em plus 0.5em minus 0.4em\relax New York, New York: Academic Press, 1975.

\bibitem{IFO}
J.~W. Goodman, \emph{Introduction to Fourier Optics}, 4th~ed.\hskip 1em plus 0.5em minus 0.4em\relax New York, New York: W. H. Freeman and Company, 2017.

\bibitem{Gaskill1978}
J.~D. Gaskill, \emph{Linear Systems, Fourier Transforms, and Optics}.\hskip 1em plus 0.5em minus 0.4em\relax Hoboken, New Jersey: Wiley, 1978.

\bibitem{Arfken2013}
G.~B. Arfken, H.~J. Weber, and F.~E. Harris, \emph{Mathematical Methods for Physicists}, 7th~ed.\hskip 1em plus 0.5em minus 0.4em\relax Waltham, Massachusetts: Academic Press, 2013.

\bibitem{Gbur_2011}
G.~J. Gbur, \emph{Mathematical Methods for Optical Physics and Engineering}.\hskip 1em plus 0.5em minus 0.4em\relax Cambridge, United Kingdom: Cambridge University Press, 2011.

\bibitem{Winocur:83}
\BIBentryALTinterwordspacing
J.~Winocur, ``Dual-wavelength adaptive optical systems,'' \emph{Appl. Opt.}, vol.~22, no.~23, pp. 3711--3715, Dec 1983. [Online]. Available: \url{https://opg.optica.org/ao/abstract.cfm?URI=ao-22-23-3711}
\BIBentrySTDinterwordspacing

\bibitem{doi:10.1088/0959-7174/2/3/003}
R.~G. Lane, A.~Glindemann, and J.~C. Dainty, ``Simulation of a {Kolmogorov} phase screen,'' \emph{Waves Random Media}, vol.~2, no.~3, pp. 209--224, 1992.

\bibitem{Frehlich:00}
R.~Frehlich, ``Simulation of laser propagation in a turbulent atmosphere,'' \emph{Appl. Opt.}, vol.~39, no.~3, pp. 393--397, Jan 2000.

\bibitem{Schmidt:10}
J.~D. Schmidt, \emph{Numerical Simulation of Optical Wave Propagation with Examples in MATLAB}.\hskip 1em plus 0.5em minus 0.4em\relax Bellingham, Washington: SPIE Press, 2010.

\end{thebibliography}
%

%

\begin{IEEEbiography}[{\includegraphics[width=1in,height=1.25in,clip,keepaspectratio]{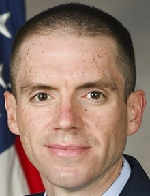}}]{Milo W. Hyde IV} (S'10--M'10--SM'12) received the B.S. degree in computer engineering from the Georgia Institute of Technology, Atlanta, GA in 2001 and the M.S. and Ph.D. degrees in electrical engineering from the Air Force Institute of Technology, Wright-Patterson Air Force Base, Dayton, OH in 2006 and 2010, respectively.

In his 23-year United States Air Force (USAF) military career, Dr. Hyde worked as a maintenance officer on the F-117A Nighthawk, as a government researcher/engineer at the Air Force Research Laboratory, as the USAF Deputy for Operations for the Defense Science Board, and finally, as a professor of electrical engineering and optical physics at the Air Force Institute of Technology.  He is the author of the book \emph{Computational Optical Coherence and Statistical Optics} and has over 150 journal and conference publications in electromagnetic material characterization, guided-wave theory, and statistical optics.

Dr. Hyde is a member of the Directed Energy Professional Society (DEPS), Sigma Xi, and a senior member of IEEE, SPIE, and OSA.
\end{IEEEbiography}

\vspace{11pt}
\begin{IEEEbiography}[{\includegraphics[width=1in,height=1.25in,clip,keepaspectratio]{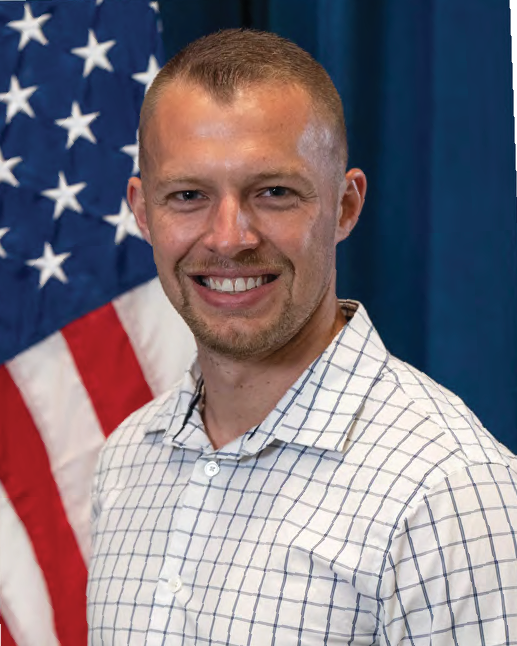}}]{Matthew Kalensky} is an engineer at the Naval Surface Warfare Center Dahlgren Division. Dr. Kalensky received his BS degree in mechanical and materials engineering from Loyola University Maryland in 2017 and received his MS and PhD in aerospace engineering from the University of Notre Dame in 2020 and 2022, respectively. Dr. Kalensky actively runs a cross-service AO working group, is a chair for the Unconventional Imaging, Sensing, and Adaptive Optics conference at SPIE Optics and Photonics, and is on the program committee for Optica’s pcAOP conference. He is also an active member of DEPS, SPIE, and Optica. Dr. Kalensky’s research interests are in beam control, atmospheric propagation, deep-turbulence characterization, and aero effects. 
\end{IEEEbiography} 

\vspace{11pt}

\begin{IEEEbiography}[{\includegraphics[width=1in,height=1.25in,clip,keepaspectratio]{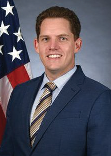}}]{Mark F. Spencer}is the Acting Principal Director for Directed Energy within the Office of the Under Secretary of Defense for Research and Engineering. Mark is also an Adjunct Associate Professor of Optical Sciences and Engineering at the Air Force Institute of Technology within the Department of Engineering Physics. He is an active member of the Directed Energy Professional Society, a senior member of Optica (the society advancing optics and photonics worldwide), and a fellow of SPIE (the international society for optics and photonics).

\end{IEEEbiography}




\end{document}